\UseRawInputEncoding
\documentclass[aps,prd,reprint,superscriptaddress]{revtex4-2}
\usepackage{geometry}
\usepackage{graphicx}
\usepackage{float}
\usepackage{subfigure}
\usepackage{slashed}
\usepackage{amsmath}
\usepackage{simplewick}   
\usepackage{tikz-feynman,contour} 
\usepackage{tikz-feynhand}
\usepackage{fancyhdr} 
\usepackage{appendix} 
\usepackage{color}
\usepackage{amssymb} 

\begin{document}

\title{A Flavor Change Study based on Dyson-Schwinger Equation}
\author{Xue-ao Chao}
\affiliation{Department of Physics and State Key Laboratory of Nuclear Physics and Technology\\
Peking University, Beijing 100871, China}
\author{Yu-xin Liu}
\email{yxliu@pku.edu.cn}
\affiliation{Department of Physics and State Key Laboratory of Nuclear Physics and Technology\\
Peking University, Beijing 100871, China}
\affiliation{Collaborative Innovation Center of Quantum Matter, Beijing 100871, China}
\affiliation{Center for High Energy Physics, Peking University, Beijing 100871, China}
\date{\today}

\newcommand{\RNum}[1]{\uppercase\expandafter{\romannumeral #1\relax}}

\begin{abstract}
    We study the flavor change effects using the Dyson-Schwinger (DS) equation in a multi-flavor system. By taking the Electroweak interaction as perturbation into conditions, the $\text{SU(4)}_\text{L}\rightarrow\text{SU(2)}_\text{L}\otimes\text{SU(2)}_\text{L}\rightarrow\text{SO(2)}_\text{L}\otimes\text{SO(2)}_\text{L}$ symmetry breaking chain is studied. Under this symmetry breaking pattern fermion masses are split, and we can identify the fermions with different masses as different generations. Quark mass spectrum is then given. Meanwhile, there are a total of fifteen Goldstones but only four of them are independent. The Goldstones have electric charges $0$, $0$, $+1$, $-1$, respectively. One of them becomes pseudo-Nambu-Goldstone boson (pNGB) and gains a mass due to the Electroweak interaction perturbation. It can be identified as the Higgs boson. The other three Goldstones maintain massless and will be eaten by gauge bosons to give $W^\pm$ and $Z^0$ masses via Schwinger mechanism. Thus, the Goldstones can take the role of the Higgs boson.
\end{abstract}

\maketitle

\section{INTRODUCTION}
    The Standard Model (SM) has been proved a great success.
    Nevertheless, it is not all the stories.
    There are many phenomena beyond the explanation given by the SM, for instance, Dark Matter, Dark Energy, and Neutrino Masses. 
    Moreover, even within the SM some problems are still puzzling scientists deeply, such as the Fine-Tuning Problem, Hierarchy Problem, Mass Splitting, and various problems related to flavors.

    Concerning flavor questions, there are four main issues 
    (The significance of these issues is present in the fact that 12 of 19 parameters in the SM are related to them. They are $m_i$ and $\Theta_k$) \cite{FAHS}:\\
    \textcircled{1} Replication: Why are there three generations of fermions?
    Each fermion particle seems to have three different versions, 
    differentiated only by mass. The three generations are literally copy-paste of the first generation. And curiously, it is pasted three times. Why three generations? What's the necessity of three generations? There is no doubt this is an important issue. The fact of three generations of fermions is vital to the cosmic structure even to the existence of life \cite{WTG}.\\
    \textcircled{2} Fermions mass hierarchy: Why so significant the mass differences between the three generations are?
    The tau lepton is roughly 3600 times more massive than the electron, and the top quark is nearly 100,000 times heavier than the up quark. Although we do not yet know the exact mass of neutrinos, related experiments have shown that they have different masses \cite{neutrino1,neutrino2,neutrino3}, and there seems to be a hierarchical structure between them. The mass spectrum is approximately exponential \cite{Blumhofer_1997}.
    The fermions mass hierarchy means that higher generation fermions will decay to the first generation quickly.
    All the matter in our daily life consists of the first generation particles, and those are protons, neutrons, and electrons. 
    The higher generation fermions only appear in high energy environments, such as cosmic rays and higher energy particle experiments. It seems to go back to our first question, what is the necessity for the existence of higher generation quarks?\\
    \textcircled{3} Mixing hierarchy: the Cabibbo-Kobayashi-Maskawa (CKM) matrix elements have hierarchies, $|V_{ii}|\sim1>|V_{12}|>|V_{23}|>|V_{13}|$.\\
    \textcircled{4} The origin of CKM and its evaluations.
    
    In this article, we mainly focus on the first two issues, trying to answer the question of why the mass of three generation quarks split. On one hand, there is not only a horizontal hierarchical structure between the three generations of fermions but also a vertical hierarchical structure between them. The horizontal hierarchies can be approximately expressed as $m_u/m_c\sim m_c/m_t\sim\lambda^4$ and $m_d/m_s\sim m_s/m_b\sim\lambda^2$, where $\lambda\approx 0.22$. If we look vertically, we can also roughly see a hierarchical structure among the masses of quarks, charged leptons, and neutrinos.
    For instance, in the first generation fermions, up and down quark have a mass of $m_u\sim m_d\sim 5\text{MeV}$, the electron has a mass of $m_e\sim 0.5\text{MeV}$, while electric neutrino has a mass of $m_1<15\text{eV}$.
    Here we are talking about current mass rather than the constituent one, which has the same origin as leptons, i.e., the Higgs mechanism. On the other hand, we notice that quarks participate in all strong interactions, electromagnetic interactions, and weak interactions. It has the strongest interaction. Electron participates in electromagnetic interaction and weak interaction, and its interaction is weaker. The electron neutrino only participates in the weakest weak interaction. The hierarchical structure of these interactions combined with the hierarchical structure of their mass suggests that their mass is likely to have a dynamic origin. In other words, the Higgs mechanism may have a dynamic mechanism, and the Higgs boson should be composite (see, e.g., Refs.~\cite{CH1,CH2,CH3,CH4}). In order to ensure the hierarchical structure of quarks, charged leptons, and neutrinos that match the strengths of the corresponding interactions, the Higgs boson needs to participate in all the three kind interactions, so it can only be composed of quarks. Based on the above arguments, we need a theory that describes consistently these two hierarchical structures. So we start by trying to take quantum chromodynamics (QCD) to explain the mass splittings of quarks. This article is the first step in this direction.

    As mentioned before, the three generation fermions with the same charge are identical if there are no mass differences. 
    The fact that mass does not affect interaction type suggests naturally that they can be treated as the same particles on the interaction level. 
    At least they can be replaced each other in interactions.
    This implies that there may be a symmetry among the three generations which is usually called horizontal symmetry \cite{FAHS,HSFTSFP}.
    Different generation fermions acquire different masses in the symmetry breaking progress.
    An up quark can be transformed to other particles in some progress, and the opposite progress occurs too.
    Just imagine when a particle (or a group of particles) transforms to an up quark, it may not distinguish between an up quark or a charm quark because they look the same.
    As a result, an up quark (first generation) can transform to a charm quark (another generation).
    Although a pure type of this flavor change progress can happen dynamically, it is forbidden kinematically to consider necessary quantum number conservation.
    In other words, a flavor change progress can not be external legs in Feynman diagram, and corresponding particles can not be regarded as stable particles \cite{FMARIAPT}.
    Even so, flavor change progress still possibly has multi-faceted impacts on many issues, for instance, CKM and mass splitting.
    Flavor change progress and its renormalization procedure have been discussed in many works (see, e.g., Refs.~\cite{FMARIAPT,FMGAWR,ET}). 

    DS equation is an important tool to study dynamical symmetry breaking \cite{DSEATATHP}.
    We will give a model from an $\text{SU(2)}_\text{L}$ horizontal symmetry, which will be dynamically broken to $\text{SO(2)}_\text{L}$ (We ignore the right-handed symmetries in this article.).
    By the tool of DS equation, we will see that there can be dynamical flavor change progress that generates different masses for different generations.
    Three generations are also indicated.

    The remainder of this article is organized as follows: In Sec. \RNum{2}, we describe the theoretical framework.
    In Sec. \RNum{3}, we discuss the structure of the solutions and the numerical results. 
    The mass spectrum for different generation quarks is obtained, and we will show that why there are three generation fermions.
    In Sec. \RNum{4}, we will discuss the goldstone physics of our model.
    Sec. \RNum{5} gives a summary and some brief remarks.

\section{THEORETICAL FRAMEWORK}
\subsection{Lagrangian and Symmetry}
    We define two types hyperquarks first, the up-type hyperquarks and the down-type hyperquarks.
    \begin{equation}\label{L1}
        \begin{aligned}
        up-type:\qquad &\left(u_1\quad u_2\right),\\
        down-type:\qquad &\left(d_1\quad d_2\right).\\
        \end{aligned}
    \end{equation}
    We consider only QCD now first; the Lagrangian density is given as:
    \begin{equation}
        L_0=\bar{u}_1i\slashed{\partial}u_1+\bar{u}_2i\slashed{\partial}u_2+\bar{d}_1i\slashed{\partial}d_1+\bar{d}_2i\slashed{\partial}d_2+L'_{QCD},
    \end{equation}
    where $L'_{QCD}$ contains the remaining terms in QCD. Obviously, it has $\text{SU(4)}_\text{L}$ invariance.
    We will add the $\text{SU(2)}_\text{L}\otimes\text{U(1)}_\text{Y}$ Electro-Weak terms as perturbation into the Lagrangian later.
    Apparently, $L_0+\text{U(1)}_\text{Y}$ terms are also invariant in the $\text{SU(4)}_\text{L}$ transformation. The symmetry breaking is due to the $\text{SU(2)}_L$ terms completely.
    We can rearrange the $\text{SU(2)}_\text{L}\otimes\text{U(1)}_\text{Y}$ terms as $\text{SU(2)}_\text{W}\otimes\text{U(1)}_\text{EM}$.
    The $\text{U(1)}_\text{EM}$ terms break the $\text{SU(4)}_L$ symmetry to $\text{SU(2)}_\text{H}\otimes\text{SU(2)}_\text{H}$ because of electric charge conservation. The subscript H suggests that the $\text{SU(2)}_\text{L}$ is horizontal symmetries under the family $\left\{u_1,u_2\right\}$ or $\left\{d_1,d_2\right\}$.

    We focus on the up sector first.
    In order to study a flavor change problem, we adopt a formulation that is able to describe the flavor change.
    The action of one of the family mentioned above after renormalization reads:
    \begin{widetext}
        \begin{eqnarray}\label{L2}
            S&=&\int\mathrm{d}^4x\big[Z_2^{11}\bar{u}_1\left(i\slashed{\partial}-Z_m^{11}m_{11}\right)u_1+Z_2^{22}\bar{u}_2\left(i\slashed{\partial}-Z_m^{22}m_{22}\right)u_2+Z_1g\bar{u}_1\gamma^\mu A^a_\mu T_a u_1+Z_1g\bar{u}_2\gamma^\mu A^a_\mu T_a u_2\nonumber\\
            &&-Z_3\frac{1}{4}F_{\mu\nu}F^{\mu\nu}-\frac{Z_3}{2\xi}\left(\partial_\mu A^\mu_a\right)^2\nonumber\\
            &&+Z_\lambda^{12}\bar{u}_1\lambda_{12}\left(i\slashed{\partial}-Z_m^{12}m_{12}\right)u_2+Z_\lambda^{21}\bar{u}_2\lambda_{21}\left(i\slashed{\partial}-Z_m^{21}m_{21}\right)u_1\nonumber\\
            &&+\frac{Z_\lambda^{12}Z_1}{Z_2}g\lambda_{12}\bar{u}_1\gamma^\mu A^a_\mu T_a u_2+\frac{Z_\lambda^{21}Z_1}{Z_2}g\lambda_{21}\bar{u}_2\gamma^\mu A^a_\mu T_a u_1\big],
        \end{eqnarray}
    \end{widetext}
    where $T_a$ are the eight Gell-Mann matrices in color space.
    The terms in the first two lines are invariant under $\text{SU(2)}_\text{L}$ transformation.
    We introduced flavor changed terms in the last two lines, which break the symmetry.
    We will let $\lambda_{ij}\rightarrow0$ at the end of calculations.
    $Z_1^{ij}$,$Z_2^{ij}$,$Z_3^{ij}$ are vertex renormalization constant, quark field, and gauge boson field renormalization constant, respectively.
    $Z_\lambda^{ij}$ is necessary for the renormalization of $\lambda$.
    Notice that we define $g=\frac{Z_2^{ii}\sqrt{Z_3}}{Z_1^{ii}}g^0$, $\lambda_{ij}=\frac{\sqrt{Z_2^{11}Z_2^{22}}}{Z_\lambda^{ij}}\lambda^0_{ij}$, where $g^0$, $\lambda^0_{ij}$ are the corresponding bare parameters, and the indexes take $\{i=1, j=2\}$ and $\{i=2, j=1\}$.
    By define a flavor doublet $\psi\equiv\left(\begin{matrix}u_1\\u_2\\ \end{matrix}\right)$, $\bar{\psi}\equiv\psi^\dagger\Gamma^0=\left(\begin{matrix}\bar{u}_1 & \bar{u}_2\\ \end{matrix}\right)$,
    where $\Gamma^\mu\equiv\left(\begin{matrix}\gamma^\mu&0\\0&\gamma^\mu\\ \end{matrix}\right)$,
    we can write the action in a compact form:
    \begin{eqnarray}
            S&=&\int\mathrm{d}^4x\big[\bar{\psi}Z_2\left(i\slashed{\partial}-M\right)\psi+\bar{\psi}gZ_1\Gamma^\mu A_\mu\psi\nonumber\\
            &&-\frac{Z_3}{4}F_{\mu\nu}F^{\mu\nu}-\frac{Z_3}{2\xi}\left(\partial_\mu A^\mu\right)^2\big],
    \end{eqnarray}
    where 
    \begin{widetext}
    \begin{eqnarray*}
        Z_1&=&\left(\begin{matrix}
            Z_1 & \frac{Z_1Z_\lambda^{12}\lambda_{12}}{Z_2}\\
            \frac{Z_1Z_\lambda^{21}\lambda_{21}}{Z_2} & Z_1\\
        \end{matrix}\right),\\
        Z_2\left(i\slashed{\partial}-M\right)&=&\left(\begin{matrix}
            Z_2^{11}\left(i\slashed{\partial}-Z_m^{11}m_{11}\right) & Z_\lambda^{12}\lambda_{12}\left(i\slashed{\partial}-Z_m^{12}m_{12}\right) \\
            Z_\lambda^{21}\lambda_{21}\left(i\slashed{\partial}-Z_m^{21}m_{21}\right) & Z_2^{22}\left(i\slashed{\partial}-Z_m^{22}m_{22}\right) \\
        \end{matrix}\right).
    \end{eqnarray*}
    \end{widetext}
    Obviously, the Lagrangian has an $\text{SU(2)}_\text{L}$ horizontal symmetry if $\lambda_{ij}=0$ and $m_{11}=m_{22}$ keep finite. In the following, we take $m_{11}=m_{22}$ indeed and let $\lambda_{ij}\rightarrow 0$ in the end. Then it will be shown that the horizontal symmetry is broken dynamically.

    For convenient numerical calculations, we switch to Euclidean space and write down the partition function as:
    \begin{eqnarray}
        Z_E\!=\!&&\int\!\!D(\bar{\psi},\psi,A_\mu)\exp\bigg\{\!\!-\!\!\int\!\mathrm{d}^4x\Big[\bar{\psi}Z_2\!\left(\slashed{\partial}+M\right)\!\psi\nonumber\\
        &&+igA_\mu\bar{\psi}Z_1\Gamma^\mu\psi+\!\frac{Z_3}{4}F_{\mu\nu}F^{\mu\nu}\!\!+\!\frac{Z_3}{2\xi}\!\left(\partial_\mu A^\mu\right)^2\!\Big]\nonumber\\
        &&+\!\!\int\!\mathrm{d}^4x\left[\bar{\psi}\eta+\bar{\eta}\psi+A_\mu J^\mu\right]\bigg\}.
    \end{eqnarray}

\subsection{DS equations}
    Following Ref.~\cite{DSEATATHP}, the DS equation for quarks is:
    \begin{eqnarray}\label{DSE}
        S^{-1}\!\left(p\right)&=&Z_2\left(i\slashed{p}+M\right)\nonumber\\
        &&+i\!\!\int\!\!\frac{\mathrm{d}^4k}{\left(2\pi\right)^4}gZ_1\Gamma^\mu S\left(k\right)\Lambda_\varepsilon  D_\mu^\varepsilon \left(p\!-\!k\right),\nonumber\\
    \end{eqnarray}
    where $S$, $D$, and $\Lambda$ stand for the propagators of quarks, gauge bosons, and the complete vertex, respectively. 
    It can be represented graphically as:
    \begin{eqnarray*}
        \begin{tikzpicture}[baseline=-0.08cm]
            \begin{feynhand}
                \vertex (p1) at (0,0);
                \vertex [dot](p2) at (1,0) {};
                \vertex [particle](p0) at (1,-0.5) {$S_0^{-1}$};
                \vertex (p3) at (2,0);
                \propag [fermion] (p1) to (p2);
                \propag [fermion] (p2) to (p3);
            \end{feynhand}
        \end{tikzpicture}^{-1}
        &=&\nonumber\\
        \begin{tikzpicture}[baseline=-0.08cm]
            \begin{feynhand}
                \vertex (p1) at (0,0);
                \setlength{\feynhandblobsize}{3mm}
                \vertex [grayblob](p2) at (1,0) {};
                \vertex [particle](p0) at (1,-0.5) {$S^{-1}$};
                \vertex (p3) at (2,0);
                \propag [fermion] (p1) to (p2);
                \propag [fermion] (p2) to (p3);
            \end{feynhand}
        \end{tikzpicture}^{-1}
        &+&
        \begin{tikzpicture}[baseline=-0.08cm]
        \begin{feynhand}
            \vertex (p0) at (0,0);
            \setlength{\feynhandblobsize}{6mm}
            \vertex [ringblob] (p1) at (1,0) {};
            \setlength{\feynhandblobsize}{3mm}
            \vertex [grayblob](p2) at (2,0){};
            \vertex [particle](p20) at (2,-0.5) {$\Sigma$};
            \vertex [dot](p3) at (3,0) {};
            \vertex (p4) at (4,0);
            \vertex [grayblob](p5) at (2,0.8) {};
            \propag [fermion] (p0) to (p1);
            \propag [fermion] (p1) to (p2);
            \propag [fermion] (p2) to (p3);
            \propag [fermion] (p3) to (p4);
            \propag [bos] (p1) to [quarter left](p5);
            \propag [bos] (p5) to [quarter left](p3);
        \end{feynhand}
        \end{tikzpicture}
    \end{eqnarray*}
    where $S_0$, $S$, and $\Sigma$ stand for the bare quark propagator, the complete quark propagator, and the self-energy of quarks, respectively.
    The gauge boson propagator is modeled by:
    \begin{equation}
        g^2D_{\mu\nu}\left(p\right)=D_{\mu\nu}^{free}p^2G\left(p\right),
    \end{equation}
    where free gauge boson propagator is
    \begin{equation}
        D_{\mu\nu}^{free}=\frac{1}{p^2}\left(g_{\mu\nu}-\frac{p_\mu p_\nu}{p^2}\right) .
    \end{equation}
    In this article, Landau gauge is in use.
    We take the bare vertex approximation and absorb $g$ into the interaction model, thus
    \begin{equation}
        \Lambda^\mu=\left(\begin{matrix}
            \gamma^\mu & 0\\
            0 & \gamma^\mu\\
        \end{matrix}\right).
    \end{equation}

    As mentioned before, we are using a formulation that can describe the flavor change progress.
    The basic idea of ​​our method to study flavor change is to regard the term that describes flavor change as a renormalization flow. The realistic version of Eq.~(\ref{L2}) is obtained by taking $\lambda_{ij}=0$. We view this as the definition of the theory at some energy scale $\xi^2$. At that scale, we can not distinguish between $S_1=S_0$ and $S_2=S_0+\lambda(\mu^2)S_\lambda$, where $S_0$ represents the first two lines in Eq.~(\ref{L2}) while $S_\lambda$ represents the other terms and the flavor change term satisfies
    \begin{equation}\label{bc}
        \lambda(\mu^2=\xi^2)=0.
    \end{equation}
    Thus, it is reasonable to assume that they describe the same theory elsewhere. When we consider the effective theory at any other scale, the structure of $S_2$ will change significantly due to radiative corrections, and $\lambda$ will generate a non-zero value. That is just the flavor change effect we desired.
    The renormalization group can be used as a dynamical tool to include all of the effects of the full theory.
    Eq.~(\ref{bc}) defines the renormalization group boundary condition for the full theory and can be viewed as our renormalization condition.
    For more related discussions, see, eg., Refs.~\cite{rg1,rg2,rg3,rg4}.

    At the renormalization point $\xi^2$, the propagator becomes free, and the interaction vertex is bare. 
    Here is our renormalization condition:
    \begin{eqnarray}\label{RC}
        &S\left(\xi^2\right)=\left(
            \begin{matrix}
                \frac{1}{i\slashed{\xi}+m_{11}} & 0 \\
                0 & \frac{1}{i\slashed{\xi}+m_{22}} \\
            \end{matrix}
        \right).
    \end{eqnarray}
    We can parameterize the quark propagator as:
    \begin{equation}
        S\!=\!\left(\!
        \begin{matrix}
            -i\slashed{p}V_1(p^2)\!+\!S_1(p^2) & -i\slashed{p}T_1(p^2)\!+\!R_1(p^2) \\
            -i\slashed{p}T_2(p^2)\!+\!R_2(p^2) & -i\slashed{p}V_2(p^2)\!+\!S_2(p^2) \\            
        \end{matrix}
        \right).
    \end{equation}
    Substituting these to Eq.~(\ref{DSE}) and carrying out many sterile derivations, we obtain eight nonlinear integral equations.
    Because these equations are too long, we only list the first equation explicitly here. The complete set of the equations are listed in appendix A.
    \begin{eqnarray}\label{NLE}
        1&=&\Big[Z_2^{11}Z_m^{11}m_{11}S_1\left(p^2\right)+Z_2^{11}p^2V_1\left(p^2\right)\nonumber\\
        &&+Z_m^{12}m_{12}R_2\left(p^2\right)+Z_\lambda^{12}\lambda_{12}p^2T_2\left(p^2\right)\Big]\nonumber\\
        &&+\int\frac{\mathrm{d}^4k}{\left(2\pi\right)^4}G(p-k)\bigg\{3\Big[S_1\left(k^2\right)S_1\left(p^2\right)\nonumber\\
        &&+R_1\left(k^2\right)R_2\left(p^2\right)\Big]+\Big[V_1\left(k^2\right)V_1\left(p^2\right)\nonumber\\
        &&+T_1\left(k^2\right)T_2\left(p^2\right)\Big]\Big[2p\cdot k\nonumber\\
        &&+\frac{\left(p^2+k^2\right)p\cdot k-2p^2k^2}{\left(p-k\right)^2}\Big]\bigg\}
    \end{eqnarray}
    All the others are structurally similar to this one.
    By the reason of simplicity and solubility, we take the point interaction model first, where $G\left(p\right)=G$ is a constant.
    Then due to the rotational symmetry of the system, we can write the 4-dimensional integration as
    \begin{equation}
        \int\frac{\mathrm{d}^4k}{(2\pi)^4}=\frac{1}{8\pi^3}\int_0^{\Lambda^2}\mathrm{d}k^2k^2\int_0^\pi\mathrm{d}\theta\sin^2\theta,
    \end{equation}
    where $\Lambda$ is the regularization momentum truncation in the numerical calculations.
    Looking closely at these formulas, we find that there is only one type of term involving angle in these integrations. That is the term like
    $$2pk\cos\theta+\frac{\left(p^2\!+\!k^2\right)pk\cos\theta\!-\!2p^2k^2}{p^2\!+\!k^2\!-\!2pk\cos\theta}\equiv\!f\left(p,k,\theta\right).$$
    Thus, using the integration identity
    \begin{equation}
        \int_{0}^{\pi}\mathrm{d}\theta \sin^2\theta f\left(p,k,\theta\right)=-\frac{3\pi}{4}p^2+\frac{\pi}{4}\frac{p^4}{k^2},
    \end{equation}
    we can integrate the angle integration out first.
    We define
    \begin{equation}\label{CVCT}
        \left\{
        \begin{aligned}
            \frac{G}{\left(8\pi\right)^3}\int\mathrm{d}k^2k^2V_1\left(k^2\right)&=CV_{11},\\
            \frac{G}{\left(8\pi\right)^3}\int\mathrm{d}k^2k^2\frac{1}{k^2}V_1\left(k^2\right)&=CV_{12},\\
            \frac{G}{\left(8\pi\right)^3}\int\mathrm{d}k^2k^2S_1\left(k^2\right)&=CS_{11},\\
            \frac{G}{\left(8\pi\right)^3}\int\mathrm{d}k^2k^2\frac{1}{k^2}S_1\left(k^2\right)&=CS_{12},\\
        \end{aligned}
        \right.
        \end{equation}
    and likewise for $V_2$, $S_2$, $T_1$, $R_1$, $T_2$, $R_2$. We have then sixteen parameters: $CV_{ij}$, $CS_{ij}$, $CT_{ij}$, and $CR_{ij}$, which are functionals of $V_i$, $S_i$, $T_i$, or $R_i$, and are independent of $p$ and $k$. As can be seen, functions of $p$ and $k$ in these equations are decoupled so that by defining the above functionals, we can collect the integrations to these parameters.
    The Eq.~(\ref{NLE}) now reads
    \begin{equation}
    \begin{aligned}
        1=&\Big[Z_2^{11}Z_m^{11}m_{11}S_1\left(p^2\right)+Z_2^{11}p^2V_1\left(p^2\right)\\
        &+Z_m^{12}m_{12}R_2\left(p^2\right)+Z_\lambda^{12}\lambda_{12}p^2T_2\left(p^2\right)\Big]\\
        &+\frac{3\pi}{2}CS_{11}S_1(p^2)+\frac{3\pi}{2}CR_{11}R_2(p^2)\\
        &-\frac{3\pi}{4}p^2CV_{11}V_1(p^2)+\frac{\pi}{4}p^4CV_{12}V_1(p^2)\\
        &-\frac{3\pi}{4}p^2CT_{11}T_2(p^2)+\frac{\pi}{4}p^4CT_{12}T_2(p^2),\\
    \end{aligned} 
    \end{equation}
    and similarly for others (see appendix A).
    With the renormalization conditions defined at renormalization point $\xi^2$ by Eqs (\ref{RC}), we have
    \begin{equation}\label{RC2}\left\{
        \begin{aligned}
            V_1\left(\xi^2\right)&=\frac{1}{\xi^2+m_{11}^2},\\
            V_2\left(\xi^2\right)&=\frac{1}{\xi^2+m_{22}^2},\\
            T_1\left(\xi^2\right)&=0,\\
            T_2\left(\xi^2\right)&=0,\\
        \end{aligned}\qquad
        \begin{aligned}
            S_1\left(\xi^2\right)&=\frac{m_{11}}{\xi^2+m_{11}^2},\\
            S_2\left(\xi^2\right)&=\frac{m_{22}}{\xi^2+m_{22}^2},\\
            R_1\left(\xi^2\right)&=0,\\
            R_2\left(\xi^2\right)&=0.\\
        \end{aligned}\right.
        \end{equation}
    Using the above formulas, we can solve the renormalization constants, and then put them into the above formulas, and finally, we get
    \begin{equation}\left\{
    \begin{aligned}
        1&\!=\!m_{ii}S_i\!\left(p^2\right)\!+\!\left(\!1\!-\!\frac{\pi}{4}\left(\xi^2\!-\!p^2\right)\!CV_{i2}\!\right)p^2V_i\!\left(p^2\right)\\
        &\qquad\qquad\qquad-\frac{\pi}{4}\left(\xi^2\!-\!p^2\right)p^2CT_{i2}T_j\!\left(p^2\right),\\
        0&\!=\!m_{ii}R_i\!\left(p^2\right)\!+\!\left(\!1\!-\!\frac{\pi}{4}\left(\xi^2\!-\!p^2\right)\!\right)\!p^2CV_{i2}T_i\!\left(p^2\right)\\
        &\qquad\qquad\qquad-\frac{\pi}{4}\left(\xi^2\!-\!p^2\right)p^2CT_{i2}V_j\!\left(p^2\right),\\
        0&\!=\!-m_{ii}V_i\!\left(p^2\right)\!+\!\left(\!1\!-\!\frac{\pi}{4}\left(\xi^2\!-\!p^2\right)\!\right)\!CV_{i2}S_i\!\left(p^2\right)\\
        &\qquad\qquad\qquad-\frac{\pi}{4}\left(\xi^2\!-\!p^2\right)CT_{i2}R_j\!\left(p^2\right),\\
        0&\!=\!-m_{ii}T_i\!\left(p^2\right)\!+\!\left(\!1\!-\!\frac{\pi}{4}\left(\xi^2\!-\!p^2\right)\!\right)\!CV_{i2}R_i\!\left(p^2\right)\\
        &\qquad\qquad\qquad-\frac{\pi}{4}\left(\xi^2\!-\!p^2\right)CT_{i2}S_j\!\left(p^2\right),\\
    \end{aligned}\right.
    \end{equation}
    for $\left\{i=1,j=2\right\}$ and $\left\{i=2,j=1\right\}$. We have taken $\lambda=0$ and $m_{12}=m_{21}=0$ in the previous derivations.
    It is found that only $\left\{CV_{12},CV_{22},CT_{12},CT_{22}\right\}$ appear in the final equations.
    Luckily, these equations are linear for $\left\{V_i\left(p^2\right),S_i\left(p^2\right),T_i\left(p^2\right),R_i\left(p^2\right)\right\}$, and $\left\{CV_{12},CV_{22},CT_{12},CT_{22}\right\}$ are just constants which are independent of momentum so that we can solve them analytically. Even so, they are still very tedious. So we just abbreviate them as: 
    \begin{equation}\label{VSTR}
    \left\{
    \begin{aligned}
    &V_i\left(p^2;CV_{12},CV_{22},CT_{12},CT_{22}\right),\\
    &S_i\left(p^2;CV_{12},CV_{22},CT_{12},CT_{22}\right),\\
    &T_i\left(p^2;CV_{12},CV_{22},CT_{12},CT_{22}\right),\\
    &R_i\left(p^2;CV_{12},CV_{22},CT_{12},CT_{22}\right).\\
    \end{aligned}\right.
    \end{equation}
    If we substitute the functions in Eq.~(\ref{VSTR}) back to the definition Eq.~(\ref{CVCT}), we can determine the four variables numerically.
    Once this is done, we have $\left\{CV_{12},CV_{22},CT_{12},CT_{22}\right\}$ then $\left\{V_i,S_i,T_i,R_i\right\}$ and all.

    Due to the Hermiticity of the Lagrangian, we have
    \begin{equation}
        S^{-1}_{ij}(p)=\gamma^{0\dagger}S_{ij}^\dagger(p)\gamma^0,
    \end{equation}
    which has been given in Ref.~\cite{ET}.
    Thus the solutions must satisfy $T_1\left(p^2\right)=T_2\left(p^2\right)$ and $R_1\left(p^2\right)=R_2\left(p^2\right)$.
    We take these relations as conditions in the following discussion.
    Therefore, we have totally three unknown variables $\left\{CV_{12},CV_{22},CT_2\equiv CT_{12}=CT_{22}\right\}$, which can be organized in three dimensional Cartesian coordinate.

\section{NUMERICAL RESULT}
\subsection{Structure of the solutions}
\begin{figure}[!]
        \includegraphics[width=7cm]{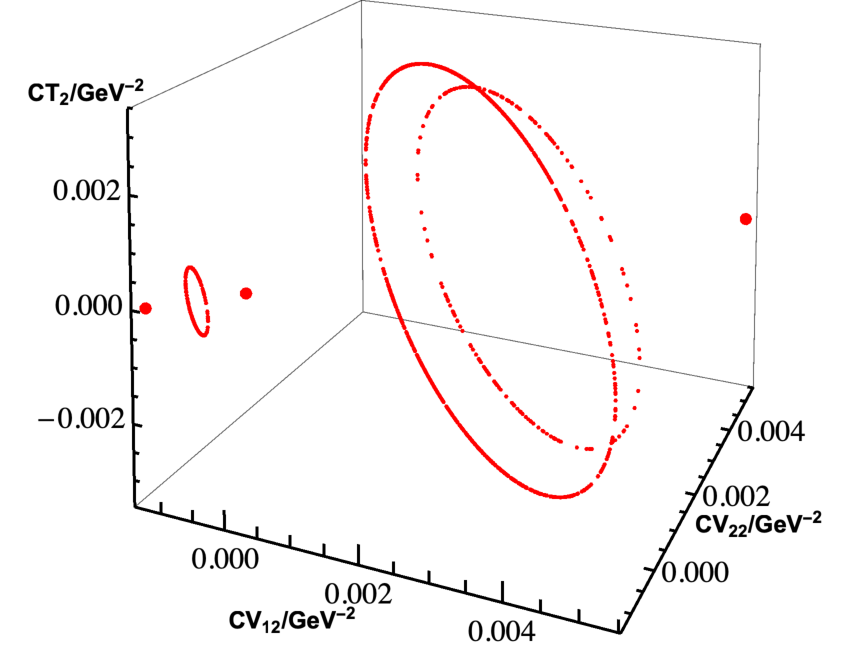}
        \caption{\label{fig1} Obtained structure of the gap solutions. These solutions are solved under the parameters as $G=1000\text{GeV}^{-2}$, $m_{11}=m_{22}=50\text{GeV}$, $\xi^2=1000\text{GV}^2$. There are two isolated (red) points on both sides of each ellipse which are the solutions that have no mixing between two flavor hyperquarks. There exist totally three such points. These points do not break the $\text{SU(2)}$ symmetry.}
    \end{figure}

    As mentioned above, one solution can be parameterized by three variables, namely $\{CV_{12},CV_{22},CT_2\}$. We can present the three variables in a three-dimensional space in which every solution is presented by a point.
    As shown in Fig.~\ref{fig1}, there are infinite solutions as expected because the original Lagrangian has a SU(2) symmetry. These solutions form exactly some ellipses. Apparently, they can be classified into three groups, and each one is an ellipse with eccentricity $e=\frac{\sqrt{2}}{2}$.
    Here is the reason why we take the point interaction model.
    In this model, we can solve the DS equations semi-analytically.
    The more important is that there are only three variables to parameterize the solutions in this model. So, it gives us an easy way to find the structure of the solutions. We can also choose other more realistic interaction models, certainly.
    However, in that case, we will be only able to solve the equations numerically, and it will be difficult to show the structure of the solutions without any advanced knowledge of it.
    
\begin{figure}[!]
        \includegraphics[width=7cm]{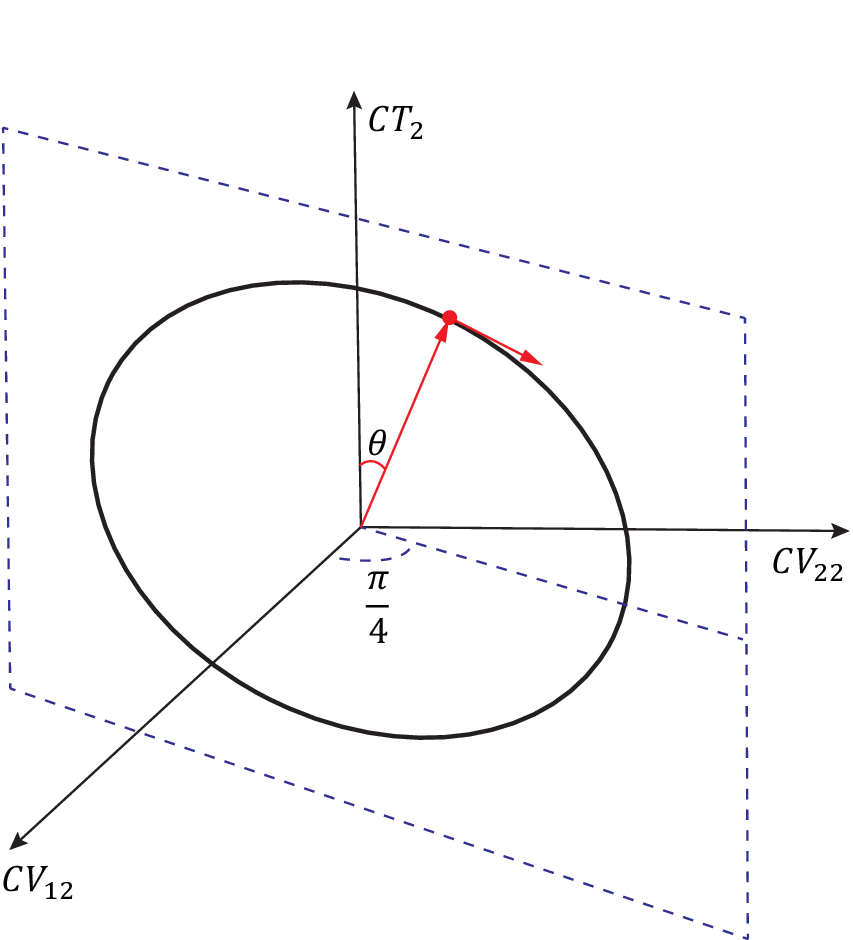}
        \caption{\label{fig2} Schematic diagram of the solutions in one ellipse.
        A rotation $\theta$ in this ellipse can be regarded as a $\text{SO(2)}$ transformation. The plane of the ellipse has an angle of $\frac{\pi}{4}$ with the $\text{CV}_{12}$ axis and $\text{CV}_{22}$ axis. Note that the origin of the coordinate system in the figure is not necessary at $(0,0,0)$.The point with $\theta=0$ ($\text{CV}_{12}=\text{CV}_{22}$) is a fixed point of $\text{SU(2)}$. The points with $\theta=\frac{\pi}{2}$ and $\theta=\frac{3\pi}{2}$ correspond to the diagonalized points.}
    \end{figure}

    On the one hand, a SO(2) rotation of $\theta'$ in the flavor space $\psi=\left(\begin{matrix}u_1\\u_2\\ \end{matrix}\right)$ can be defined by
    \begin{equation}\label{R1}
    \left(\begin{matrix}
        u'_1 \\
        u'_2 \\
    \end{matrix}\right)=
    \left(\begin{matrix}
        \cos{\theta'} & -\sin{\theta'} \\
        \sin{\theta'} & \cos{\theta'} \\
    \end{matrix}\right)
    \left(\begin{matrix}
        u_1 \\
        u_2 \\
    \end{matrix}\right).   
    \end{equation}
    It is not hard to find that it induces a rotation $\theta$ defined by
    \begin{equation}\label{R2}
        \begin{aligned}
        &\qquad\qquad\qquad\qquad
        \left(
            \begin{matrix}
                CV_{12} & CT_2\\
                CT_2 & CV_{22}\\
            \end{matrix}
        \right)\rightarrow\\
        &\left(
            \begin{matrix}
                \cos\theta & -\sin\theta\\
                \sin\theta & \cos\theta\\
            \end{matrix}
        \right)\left(
            \begin{matrix}
                CV_{12} & CT_2\\
                CT_2 & CV_{22}\\
            \end{matrix}
        \right)\left(
            \begin{matrix}
                \cos\theta & \sin\theta\\
                -\sin\theta & \cos\theta\\
            \end{matrix}
        \right)
        \end{aligned}
    \end{equation}
    in the parameter space.
    The rotation $\theta$ is also equivalent to the one in the ellipse, as showing in Fig.~\ref{fig2}.
    The relation between the two rotations is
    \begin{equation}
        \tan{2\theta'}=\frac{\tan{\theta}}{\sqrt{2}}.
    \end{equation}
    This suggests that the system still has a SO(2) symmetry.
    The other two transformation directions in the SU(2) group drive the transformed point out of the ellipse.
    However, the point $CV_{12}=CV_{22}$ ($\theta=0$ on the ellipse) is a fixed point of SU(2) translation.
    So we can say that the SU(2) symmetry is broken to SO(2) symmetry.

    On the other hand, we can diagonalize the propagator or the mass matrix by a rotation defined in Eq.~(\ref{R1}) or Eq.~(\ref{R2}) to obtain the physical states.
    The fact that the eccentricity of each ellipse is always $\frac{\sqrt{2}}{2}$ guarantees whatever point we choose in the ellipse, after the diagonalization, we get always the same result.
    This means each point in the same ellipse corresponding to the same physical state.
    Therefore, they are degenerate, and every different ellipse maps to a different physical state one by one. In other words, there is a one-to-one correspondence between the ellipses and the physical states.
    The diagonalization procedure is actually equivalent to rotate the point on the ellipse to the major axis point.
    If we interpret the different excited states of hyperquarks as different particles, each ellipse corresponds to one particle. We can then regard the states in the three ellipses as the up family quarks $\{u,c,t\}$ with the same quantum numbers except for masses.

\subsection{Mass splitting and spectrum}
\begin{figure}[!]
        \includegraphics[width=7cm]{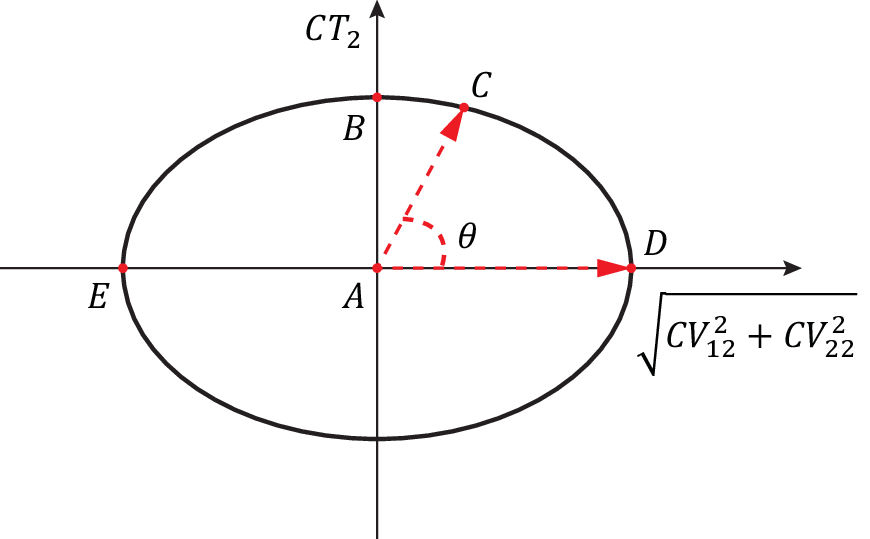}
        \caption{\label{fig3} Schematic diagram of the ellipse. The coordinates in the $(CV_{12}, CV_{22}, CT_2)$ space of characteristic points on the ellipse are as follows. A: $\left(c_0, c_0, 0\right)$, B: $\left(c_0, c_0, \frac{a_0}{\sqrt{2}}\right)$, C: $\left(CV_{12}, CV_{22}, CT_2\right)$, D: $\left(\frac{c_0-\sqrt{2}a_0}{2}, \frac{c_0+\sqrt{2}a_0}{2}, 0\right)$, E: $\left(\frac{c_0+\sqrt{2}a_0}{2}, \frac{c_0-\sqrt{2}a_0}{2}, 0\right)$.}
    \end{figure}

    As mentioned above, by diagonalizing the propagator matrix, we can get the mass spectrum of up family quarks.
    So do the down family quarks.
    The solution can be parameterized by an ellipse equation
    \begin{subequations}
        \begin{eqnarray}
            \left(\!CV_{12}-\frac{c_0}{2}\!\right)^2\!\!\!+\!\left(\!CV_{22}-\frac{c_0}{2}\!\right)^2\!\!+2CT_2^2&&=a_0^2,\qquad\\
            CV_{12}+CV_{22}&&=c_0,
        \end{eqnarray}    
    \end{subequations}
    as shown in Fig.~\ref{fig3}.
    The $a_0$ determines the size of the ellipse, and the $c_0$ determines its position.
    We only need to obtain at least two points in the ellipse, for figuring out $a_0$ and $c_0$.
    The parameter matrix 
    $
        \left(
            \begin{matrix}
                CV_{12} & CT_2 \\
                CT_2 & CV_{22} \\
            \end{matrix}
        \right)
    $
    can be diagonalized to
    \begin{equation}
        \left(
            \begin{matrix}
                \frac{c_0+\sqrt{2}a_0}{2} & 0 \\
                0 & \frac{c_0-\sqrt{2}a_0}{2} \\
            \end{matrix}
        \right).
    \end{equation}
    This corresponds to a diagonalized propagator
    \begin{equation}
        S=\left(
            \begin{matrix}
                i\slashed{p}V_1(p^2)+S_1(p^2) & 0 \\
                0 & i\slashed{p}V_2(p^2)+S_2(p^2) \\
            \end{matrix}
        \right),
    \end{equation}
    where
    \begin{equation}
    \left\{\begin{aligned}
        &V_1(p^2)=\frac{1+\frac{\pi}{4}\frac{c_0+\sqrt{2}a_0}{2}(p^2-\xi^2)}{m_{11}^2+p^2\left(1+\frac{\pi}{4}\frac{c_0+\sqrt{2}a_0}{2}(p^2-\xi^2)\right)^2},\\
        &S_1(p^2)=\frac{m_{11}}{m_{11}^2+p^2\left(1+\frac{\pi}{4}\frac{c_0+\sqrt{2}a_0}{2}(p^2-\xi^2)\right)},\\
        &M_1(p^2)=\frac{S_1(p^2)}{V_1(p^2)}=\frac{m_{11}}{1+\frac{\pi}{4}\frac{c_0+\sqrt{2}a_0}{2}(p^2-\xi^2)},
    \end{aligned}\right.
    \end{equation}
    \begin{equation}
        \left\{\begin{aligned}
            &V_2(p^2)=\frac{1+\frac{\pi}{4}\frac{c_0-\sqrt{2}a_0}{2}(p^2-\xi^2)}{m_{22}^2+p^2\left(1+\frac{\pi}{4}\frac{c_0-\sqrt{2}a_0}{2}(p^2-\xi^2)\right)^2},\\
            &S_2(p^2)=\frac{m_{22}}{m_{22}^2+p^2\left(1+\frac{\pi}{4}\frac{c_0-\sqrt{2}a_0}{2}(p^2-\xi^2)\right)},\\
            &M_2(p^2)=\frac{S_1(p^2)}{V_1(p^2)}=\frac{m_{22}}{1+\frac{\pi}{4}\frac{c_0-\sqrt{2}a_0}{2}(p^2-\xi^2)}.
        \end{aligned}\right.
        \end{equation}
    The off-diagonal terms which describes the flavor mixed effects become zero because $T_1=R_1=T_2=R_2=0$ after diagonalization.
    
\begin{figure}[!]
        \includegraphics[width=7cm]{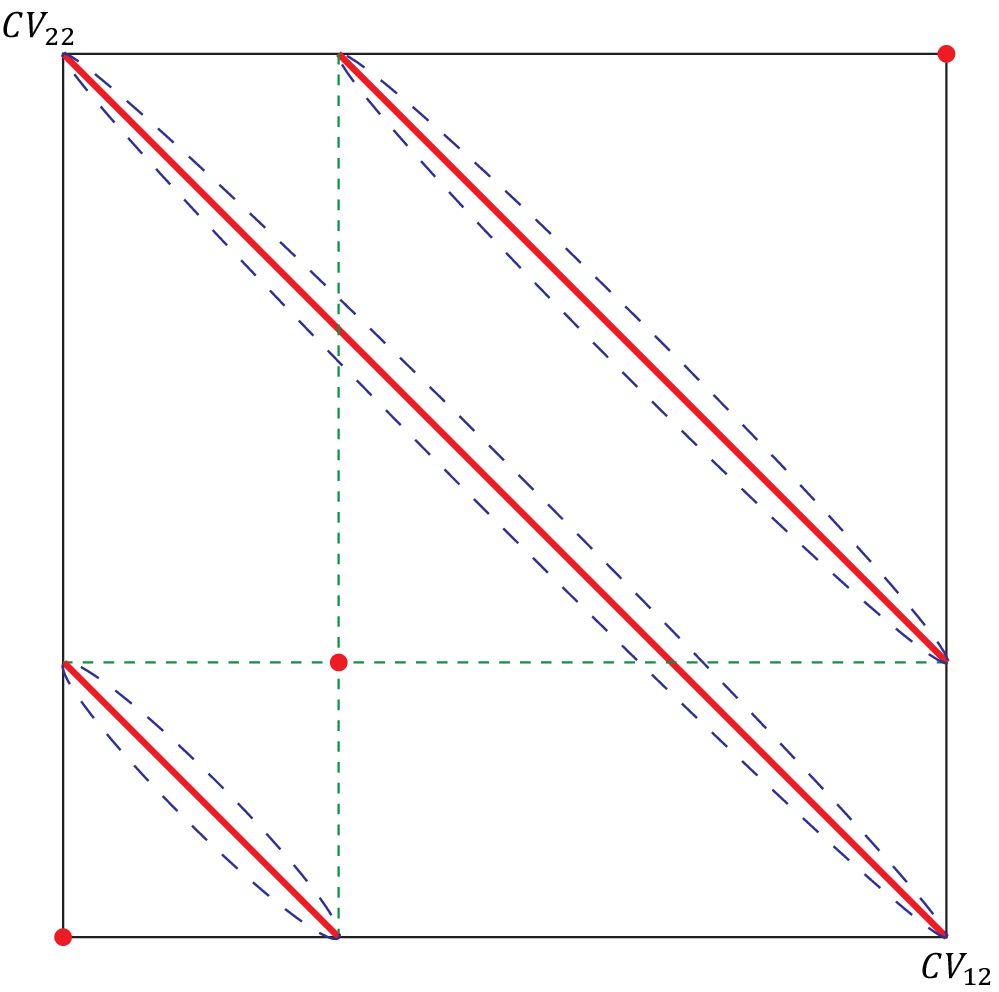}
        \caption{\label{fig4} Top view of the solutions. Each ellipse corresponds to two diagonalized states, but they are repeated in pairs. The diagonalized states are at the ellipses' major axis point, but they can also be related numerically to the red points in this figure. There are three different diagonalized states corresponding to physical states shown by red points.}
    \end{figure}

\begin{figure}[!]
        \subfigure{
        \includegraphics[width=7cm]{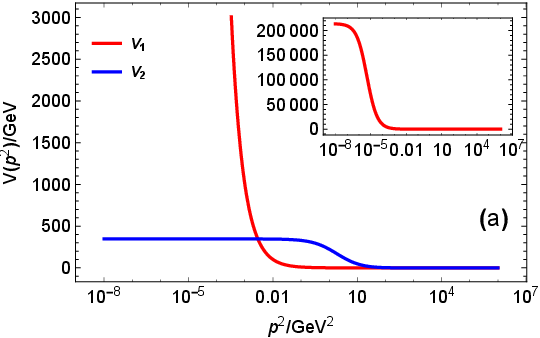}
            }
        \subfigure{
        \includegraphics[width=7cm]{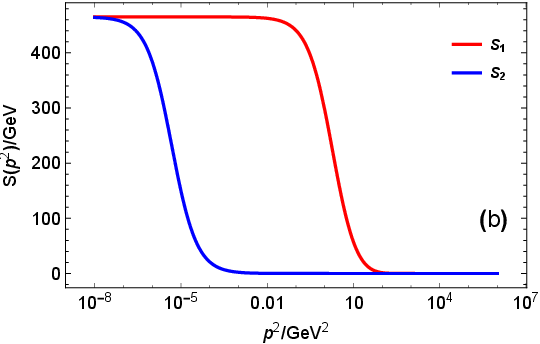}
            }
        \subfigure{
        \includegraphics[width=7cm]{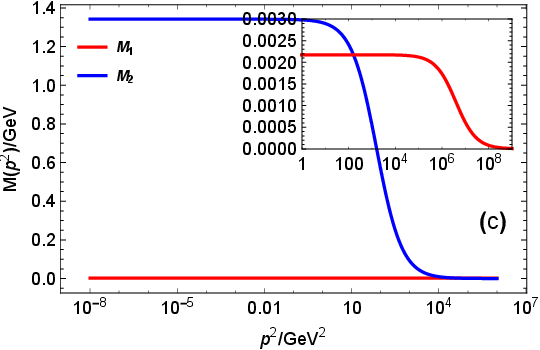}
            }
        \caption{\label{fig5} The diagonalized propagator structure functions. The parameters are taken as: $\xi^2=4.52\times 10^4 \text{GeV}^2$, $G=2.88\times 10^{-6}\text{GeV}^{-2}$, $m_{11}=m_{22}=2.15\text{MeV}$. The two color lines correspond to the two different solutions obtained by diagonalization. (a): The vector structure functions $V_i(p^2)$. (b): The scalar structure functions $S_i(p^2)$. (c): The mass functions $M_i(p^2)$.}
    \end{figure}

    There are three groups of different solutions in total and the relationships between them are quite interesting and subtle. If we look at the top view of it, these ellipses are always from the top left to the bottom right, inscribed to the same square, as shown in Fig.~\ref{fig4}. Each solution corresponds to an ellipse in the figure and can be diagonalized to obtain two different mass eigenvalues. Therefore, we can obtain six mass eigenvalues totally, but they are repeated in pairs, so we only obtain three different results finally.
    According to Ref.~\cite{Blumhofer_1997}, one may interpret either the different solutions of a gap equation or the different propagator poles of one solution as particles of different generations.
    Thus, we can obtain the three generation quark masses from the diagonalized propagator.
    When $\xi^2=4.52\times 10^4 \text{GeV}^2$, $G=2.88\times 10^{-6}\text{GeV}^{-2}$, $m_{11}=m_{22}=2.15\text{MeV}$, the three diagonalized solutions (corresponding to the three red points in Fig.~\ref{fig4}) are $(2.812\times10^{-5},2.812\times10^{-5},0)$, $(2.817\times10^{-5},2.817\times10^{-5},0)$, and $(3.223\times10^{-7},3.223\times10^{-7},0)$ in unit $\text{GeV}^{-2}$. Corresponding to one ellipse in the figure, after the diagonalization, the propagator structure functions and the mass functions under this group of parameters are shown in Fig.~\ref{fig5}. From these mass functions, we can obtain the constituent mass and the current mass. We take the mass value at 2GeV as the current mass here. However, it is worth noting that the definition of the current masses of heavy quarks is slightly different in the next section. We get then $m_u=2.17\text{MeV}$, $m_c=1.27\text{GeV}$, and $m_t=171\text{GeV}$. As we see, the mass function is large but limited in the infrared region and tends to zero in the ultraviolet region. Besides, their masses have split as we expected.

\subsection{Numerical results with a more realistic interaction model}

    While the point interaction model described above clearly shows the structure of the solutions, it has some problems.
    First of all, the mass $m_{11}=m_{22}\neq 0$ can not be set to zero because of infrared divergence.
    Secondly, the mass function may be divergent at some scale and becomes negative.
    As we can see, the denominator of the mass function is $1+\frac{\pi}{4}\frac{c_0-\sqrt{2}a_0}{2}(p^2-\xi^2)$, so if $c_0-\sqrt{2}a_0>\frac{8}{\pi}\xi^2$, the mass function will diverge at $p^2=\xi^2-\frac{8}{\pi\left(c_0-\sqrt{2}a_0\right)}$ and becomes negative below it.
    Lastly, higher generation quarks always have more masses while lower generation quarks have fewer masses at the same momentum scale. Clearly, the mass functions shown in Fig.~\ref{fig5} do not have correct running behaviors.

\begin{figure}[!]
        \subfigure{
        \includegraphics[width=7cm]{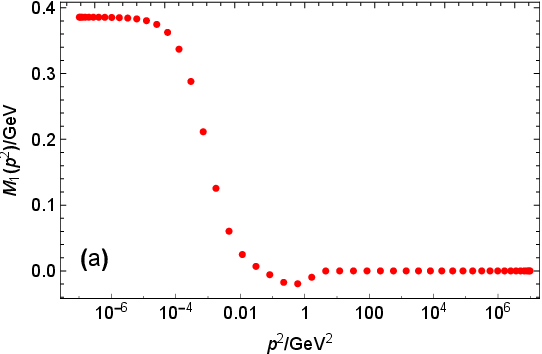}
        }
        \subfigure{
        \includegraphics[width=7cm]{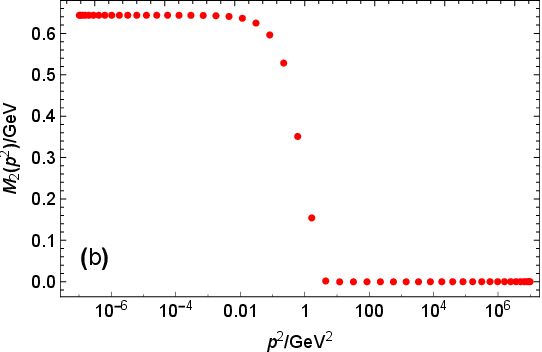}
        }
        \subfigure{
        \includegraphics[width=7cm]{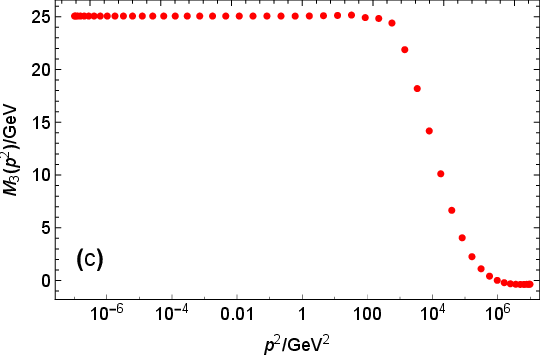}
        }
        \caption{\label{fig7} The obtained mass function in terms of the $p^2$. The parameters $e^2=21$, $\mu^2=1000\text{GeV}^2$, $\xi^2=10^6\text{GeV}^2$, $D=0.9\text{GeV}^2$, $\omega=0.4\text{GeV}$, $m_{11}=m_{22}=0\text{GeV}$ are taken. (a): The mass function of the lightest quark, $m_1$ or $m_u$. (b): The mass function of the intermediate mass quark, $m_2$ or $m_c$. (c): The mass function of the heaviest quark, $m_3$ or $m_t$.}
    \end{figure}

    These problems arise from the inappropriate interaction model. In our point interaction model, the interaction function $G(p)$ is constant while the dominant strong interaction is asymptotically free actually. So we should take a more realistic interaction model.
    Different interaction models are embodied in different $G(p^2)$.
    We take
    \begin{equation}
        G\left(p^2\right)=\frac{4}{3}G_1\left(p^2\right)+G_2\left(p^2\right),        
    \end{equation}
    where $\frac{4}{3}$ arises from the sum of color matrices and
    \begin{equation}
        G_1\left(p^2\right)=\frac{8\pi^2}{\omega^4}De^{-\frac{p^2}{\omega^2}}
    \end{equation}
    is the infrared constant model (usually simply say QC model) \cite{QCModel} in order to describe the strong interaction, while
    \begin{equation}
        G_2\left(p^2\right)=\frac{e^2}{p^2+\mu^2},
    \end{equation}
    in order to model the other remainder interactions including electromagnetic interaction.
    Because we are using a more complex model, we are only able to carry out pure numerical calculations now. 
    Thanks to the previous point interaction model, we have understood the basic structure of the solutions, which makes the numerical calculation less difficult.
    Using the same method as before, we can deal with the current numerical problems only needing a simple expansion.
    It is proved that the above discussions on the basic structures of the solutions are also valid under the current model. 
    Under a group of typical parameters, an example of the obtained mass functions are shown in Fig.~\ref{fig7}.
    The best fitted results are shown in Table \ref{tab1}.
    \begin{table}[!]
        \caption{\label{tab1} Parameter $e^2$ dependence of the best fitted constituent mass spectrum. See the footnote for the definition of quality in parentheses. $\xi^2=10^6\text{GeV}^2$, $\mu^2=1000\text{GeV}^2$, $D=0.9\text{GeV}^2$, $\omega=0.4\text{GeV}$, $m_{11}=m_{22}=0\text{GeV}$. Dimentional quantities are in unit GeV.}
        \begin{ruledtabular}
        \begin{tabular}{cccc}
            $e^2$ & $m_1$ & $m_2$ & $m_3$ \\
            \hline
            $33$ & $0.48$\footnote{The current mass is not obtained because of the negative value at 2GeV; refer to (a) in Fig.~\ref{fig7}.} & $0.95(0.77)$\footnote{Mass at the pole.} & $166(137)$\footnotemark[2] \\
            $20$ & $0.51(0.00080)$\footnote{Mass at 2GeV.} & $0.63(0.016)$\footnotemark[3] & $14(13)$\footnotemark[2] \\
        \end{tabular}
        \end{ruledtabular}
    \end{table}
    We define the value of the mass function at the zero point as the constituent mass of the quark. The current masses of heavy quarks and light quarks adopt different definitions (see the footnotes in the table).

    It can be seen that the masses of different generation quarks have obvious split, and the higher generation the quarks are, the more massive the quarks are. The constituent masses of all flavor quarks are greater than their current masses. The constituent masses of heavy flavor quarks are almost the same as their current masses, while the constituent masses of light flavor quarks are significantly different from their current masses.
    
\section{Goldstone Bosons}

    Now let us take the down-type sector into account and discuss Goldstone physics. The $\text{SU(2)}_\text{L}$ symmetry of the down-type hyperquarks is also dynamically broken to $\text{SO(2)}_\text{L}$.
    We assume that the $\text{SU(4)}_\text{L}$ symmetry is also dynamically (or spontaneously) broken to $\text{SU(2)}_\text{L}\otimes\text{SU(2)}_\text{L}$ at some higher scale due to something more fundamental.
    The $U(1)_{EM}$ terms may be one of the candidate reasons.
    As we said at the beginning of the second section, the $\text{U(1)}_\text{EM}$ terms break the $\text{SU(4)}_\text{L}$ symmetry to $\text{SU(2)}_\text{H}\otimes\text{SU(2)}_\text{H}$ because of electric charge conservation. The $\text{U(1)}_\text{EM}$ terms are the mixing of the $\text{U(1)}_\text{Y}$ and $\text{SU(2)}_\text{L}$.
    Given that the $\text{U(1)}_\text{Y}$ terms are invariant under the $\text{SU(4)}_L$ symmetry and we regard the $\text{SU(2)}_\text{L}$ terms as perturbative, it appears that $\text{SU(4)}_L$ is spontaneously broken because the original Lagrangian is $\text{SU(4)}_L$ invariant. 
    It remains to be shown whether this argument makes sense or not. However, we will not address to that question here.
    The symmetry breaking pattern is $\text{SU(4)}_\text{L}\rightarrow\text{SO(2)}_\text{L}\otimes\text{SO(2)}_\text{L}$ now. So there will be $15-2=13$ Goldstone bosons totally.
    We follow the approach provided in Ref.~\cite{NGBCBTOPISSB} to analyze these Goldstones. Let us briefly review the approach here first.

    Introducing the symmetry breaking term to the Lagrangian as
    \begin{equation}
        L_{SB}=\lambda\bar{\psi}M\psi,
    \end{equation}
    where $M\in g=\{T_a\}$, and the $T_a$ are the basis of the Lie algebra of the symmetry group,
    one can define the order parameter as
    \begin{equation}
        \Delta=\left\langle\bar{\psi}M\psi\right\rangle.
    \end{equation}
    In the limit $\lambda\rightarrow0$, if the order parameter $\Delta$ is not zero, then the corresponding symmetry generated by $T_a$ with $[T_a,M]\neq0$ is broken.
    The corresponding susceptibility is defined as 
    \begin{equation}
        \chi_{\Delta}=\lim\limits_{\lambda\rightarrow0}\frac{\Delta}{\lambda}.
    \end{equation}
    Then, the divergence of $\chi_\Delta$ suggests a spontaneous symmetry breaking.
    The Goldstones can be characterized by 
    \begin{equation}\label{GB}
        \Pi_a=i\bar{\psi}\left[T_a,M\right]\psi.
    \end{equation} 
    
    Here, we choose the matrix $M=T_1+T_{13}$, where $T_i$, $i=1,\cdots,15$, are the generators of SU(4).
    We define
    \begin{eqnarray}
        \Delta_1=&&\langle\bar{\psi}T_1\psi\rangle,\\
        \Delta_2=&&\langle\bar{\psi}T_{13}\psi\rangle,
    \end{eqnarray}
    then, it can be shown that
    \begin{equation}
        \sum_cf_{ac1}^2\Delta_1+\sum_cf_{ac13}^2\Delta_2=\lambda D_{aa}(\omega=0,\textbf{q}=0),
    \end{equation}
    where $D_{ij}$ stand for propagators of Goldstones.
    This formula is an extended multi-flavor version of what Takashi Yanagisawa obtained there. See appendix B for the proof.
    It can be seen that the divergence of susceptibility means $\Pi_a$ has a pole at $\omega=0$ and $q=0$, so $\Pi_a$ is indeed massless.
    The symmetry breaking term is then
    \begin{equation}
        L_{SB}=\lambda\!
        \left(\begin{matrix}
            \bar{u}_1 & \bar{u}_2 & \bar{d}_1 & \bar{d}_2 
        \end{matrix}\right)_{\!R}\!\!
        \left(\begin{matrix}
            0 & 1 & 0 & 0 \\
            1 & 0 & 0 & 0 \\
            0 & 0 & 0 & 1 \\
            0 & 0 & 1 & 0 \\ 
        \end{matrix}\right)\!\!
        \left(\begin{matrix}
            u_1 \\
            u_2 \\
            d_1 \\
            d_2 \\
        \end{matrix}\right)_{\!\!L},
    \end{equation}
    and the order parameter is 
    \begin{equation}
        \Delta=\left\langle\bar{\psi}M\psi\right\rangle=\left\langle\bar{u}_{2_R}u_{1_L}+\bar{d}_{2_R}d_{1_L}\right\rangle+h.c.
    \end{equation}
    When there is flavor change between $\{u_1,u_2\}$ and $\{d_1,d_2\}$, the symmetry is broken.
    The corresponding symmetry breaking pattern is desired as $\text{SU(4)}_L\rightarrow\text{SO(2)}_L\otimes\text{SO(2)}_L$.

    According to Eq.~(\ref{GB}), the Goldstones can be written exactly as (For brevity, we omit the subscripts L and R in the following formula, but keep that in mind.):
    \begin{equation}
    \begin{aligned}
        &\pi_1=0,\\
        &\pi_2=2\left(\bar{u}_1u_1-\bar{u}_2u_2\right),\\
        &\pi_3=2i\left(\bar{u}_1u_2-\bar{u}_2u_1\right),\\
        &\pi_4=i\left(\bar{u}_1d_2-\bar{u}_2d_1+\bar{d}_1u_2-\bar{d}_2u_1\right),\\
        &\pi_5=\left(\bar{u}_1d_2-\bar{u}_2d_1-\bar{d}_1u_2+\bar{d}_2u_1\right),\\
        &\pi_6=i\left(\bar{u}_2d_2-\bar{u}_1d_1+\bar{d}_1u_1-\bar{d}_2u_2\right),\\
        &\pi_7=\left(\bar{u}_2d_2-\bar{u}_1d_1-\bar{d}_1u_1+\bar{d}_2u_2\right),\\
        &\pi_8=\frac{2i}{\sqrt{3}}\left(\bar{d}_2d_1-\bar{d}_1d_2\right),\\
        &\pi_9=i\left(\bar{u}_1d_1-\bar{u}_2d_2+\bar{d}_2u_2-\bar{d}_1u_1\right),\\
        &\pi_{10}=\left(\bar{u}_1d_1-\bar{u}_2d_2-\bar{d}_2u_2+\bar{d}_1u_1\right),\\
        &\pi_{11}=i\left(\bar{u}_2d_1-\bar{u}_1d_2+\bar{d}_2u_1-\bar{d}_1u_2\right),\\
        &\pi_{12}=\left(\bar{u}_2d_1-\bar{u}_1d_2-\bar{d}_2u_1+\bar{d}_1u_2\right),\\
        &\pi_{13}=0,\\
        &\pi_{14}=2\left(\bar{d}_1d_1-\bar{d}_2d_2\right),\\
        &\pi_{15}=\left(\sqrt{\frac{3}{2}}+\sqrt{\frac{1}{6}}\right)i\left(\bar{d}_1d_2-\bar{d}_2d_1\right).\\
    \end{aligned}
    \end{equation}
    As we can see, they are not all completely independent. There are only eight independent Goldstones. After appropriate linear combinations, we can identify them as (ignoring the normalization constant):
    \begin{equation}
        \begin{array}{ll}
            \Pi_1=\pi_2, &\qquad\Pi_5=\pi_3,\\
            \Pi_2=\pi_{14}, &\qquad\Pi_6=\pi_{15},\\
            \Pi_3=\pi_7-i\pi_6, &\qquad\Pi_7=i\pi_5+\pi_4,\\
            \Pi_4=\pi_7+i\pi_6, &\qquad\Pi_8=i\pi_5-\pi_4.\\
        \end{array}
    \end{equation}
    It is obvious that $\Pi_3$ conjugate with $\Pi_4$ and $\Pi_7$ conjugate with $\Pi_8$ (ignoring the unimportant overall phase).

    Under the preserved transformation $M=T_1+T_{13}$, the hyperquarks transform as:
    \begin{equation}
    \begin{aligned}
        &\left(\begin{matrix}
            u_1\\
            u_2\\
        \end{matrix}\right)_L
        \rightarrow
        \left(\begin{matrix}
            \cos{\theta_1}&i\sin{\theta_1}\\
            i\sin{\theta_1}&\cos{\theta_1}\\
        \end{matrix}\right)
        \left(\begin{matrix}
            u_1\\
            u_2\\
        \end{matrix}\right)_L,\\
        &    \left(\begin{matrix}
            d_1\\
            d_2\\
        \end{matrix}\right)_L
        \rightarrow
        \left(\begin{matrix}
            \cos{\theta_2}&i\sin{\theta_2}\\
            i\sin{\theta_2}&\cos{\theta_2}\\
        \end{matrix}\right)
        \left(\begin{matrix}
            d_1\\
            d_2\\
        \end{matrix}\right)_L.\\
    \end{aligned}
    \end{equation}
    So, the Goldstones transform as
    \begin{equation}
    \begin{aligned}
        \Pi_5&=i\left(\bar{u}_{1_R}u_{2_L}-\bar{u}_{2_R}u_{1_L}\right)\\
        &=
        \left(\begin{matrix}
            \bar{u}_1 & \bar{u}_2 \\
        \end{matrix}\right)_R
        \left(\begin{matrix}
            0 & i \\
            -i & 0 \\
        \end{matrix}\right)
        \left(\begin{matrix}
            u_1 \\ u_2 \\
        \end{matrix}\right)_L\\
        &\rightarrow
        \left(\begin{matrix}
            \bar{u}_1 & \bar{u}_2 \\
        \end{matrix}\right)_R
        \left(\begin{matrix}
            c&-is\\
            -is&c\\
        \end{matrix}\right)
        \left(\begin{matrix}
            0 & i \\
            -i & 0 \\
        \end{matrix}\right)\\
        &\qquad\qquad\qquad\quad\times\left(\begin{matrix}
            c&is\\
            is&c\\
        \end{matrix}\right)
        \left(\begin{matrix}
            u_1 \\ u_2 \\
        \end{matrix}\right)_{\!\!L}\\
        &=
        \left(\begin{matrix}
            \bar{u}_1 & \bar{u}_2 \\
        \end{matrix}\right)_R
        \left(\begin{matrix}
            -\sin{2\theta} & i\cos{2\theta} \\
            -i\cos{2\theta} & \sin{2\theta} \\
        \end{matrix}\right)
        \left(\begin{matrix}
            u_1 \\ u_2 \\
        \end{matrix}\right)_{\!\!L}.\\
    \end{aligned}
    \end{equation}
    Because of
    \begin{equation}
    \begin{aligned}
        \Pi_1&=\left(\bar{u}_{1_R}u_{1_L}-\bar{u}_{2_R}u_{2_L}\right)\\
        &=
        \left(\begin{matrix}
            \bar{u}_1 & \bar{u}_2 \\
        \end{matrix}\right)_R
        \left(\begin{matrix}
            1 & 0 \\
            0 & -1 \\
        \end{matrix}\right)
        \left(\begin{matrix}
            u_1 \\ u_2 \\
        \end{matrix}\right)_{\!\!L},\\
    \end{aligned}
    \end{equation}
    we see that $\Pi_5$ is transformed to $\Pi_1$ under the preserved symmetry group $\text{SO(2)}_L\otimes\text{SO(2)}_L$ by $\theta=-\frac{\pi}{4}$.
    Similarly, $\Pi_6$ is transformed to $\Pi_2$, $\Pi_7$ to $\Pi_3$, and $\Pi_8$ to $\Pi_4$.
    We can thus conclude that each two Goldstones are degenerate. We have, in turn, only four Goldstones to deal with.
    They are:
    \begin{equation}
        \begin{aligned}
            \Pi_1&=\bar{u}_{1_R}u_{1_L}-\bar{u}_{2_R}u_{2_L},\\
            \Pi_2&=\bar{d}_{1_R}d_{1_L}-\bar{d}_{2_R}d_{2_L},\\
            \Pi_3&=\bar{u}_{2_R}d_{2_L}-\bar{u}_{1_R}d_{1_L},\\
            \Pi_4&=\bar{d}_{2_R}u_{2_L}-\bar{d}_{1_R}u_{1_L}.\\
        \end{aligned}  
    \end{equation}

    When we consider the $\text{SU(2)}_\text{W}$ terms, the $\text{SU(4)}_L$ symmetry becomes approximate so that some of the Goldstones become pNGBs.
    Notice that $\Pi_1$ and $\Pi_2$ are due to the horizontal $\text{SU(2)}_L$ symmetry, which is exactly broken by the $\text{SU(2)}_\text{W}$ terms, so one mixing state of them must become pNGB and have a small mass while the other one is still massless.
    On the contrary, $\Pi_3$ and $\Pi_4$ are due to the vertical $\text{SU(2)}_L$ symmetry, which is not broken; therefore, they are still massless.
    Since the up-type quarks have electric charge $+\frac{2}{3}$ and down-type quarks have electric charge $-\frac{1}{3}$, $\Pi_1$ and $\Pi_2$ have charge $0$ while $\Pi_3$ has charge $-1$ and $\Pi_4$ has charge $+1$.

    According to the well-known Schwinger mechanism \cite{SchwingerMechanism1,SchwingerMechanism2,SchwingerMechanism3,SchwingerMechanism4,SchwingerMechanism5,SchwingerMechanism6}, gauge bosons can absorb a massless boson to gain their masses.
    They eat the corresponding bosons as their longitudinal component.
    This can be schematically represented as below ($\alpha$ and $\beta$ represent the relative phase between $\Pi_1$ and $\Pi_2$):
    \[
        \begin{tikzpicture}
        \begin{feynhand}
            \vertex (p1) at (0,0);
            \vertex (p2) at (1,0);
            \vertex (p3) at (2.5,0);
            \vertex (p4) at (3.5,0);
            \vertex (p5) at (5,0);
            \vertex (p6) at (6,0);
            \propag [bos] (p1) to [edge label = $Z^0$] (p2);
            \propag [bos] (p5) to [edge label = $Z^0$] (p6);
            \propag [fermion] (p2) to [half left](p3);
            \propag [fermion] (p3) to [half left](p2);
            \propag [fermion] (p4) to [half left](p5);
            \propag [fermion] (p5) to [half left](p4);
            \vertex (p31) at (2.5,0.07);
            \vertex (p32) at (2.5,-0.07);
            \vertex (p41) at (3.5,0.07);
            \vertex (p42) at (3.5,-0.07);
            \propag [fermion] (p31) to [edge label = $\alpha\Pi_1+\beta\Pi_2$] (p41);
            \propag [fermion] (p42) to (p32);
        \end{feynhand}
        \end{tikzpicture}
    \]
    \[
        \begin{tikzpicture}
        \begin{feynhand}
            \vertex (p1) at (0,0);
            \vertex (p2) at (1,0);
            \vertex (p3) at (2.5,0);
            \vertex (p4) at (3.5,0);
            \vertex (p5) at (5,0);
            \vertex (p6) at (6,0);
            \propag [bos] (p1) to [edge label = $W^-$] (p2);
            \propag [bos] (p5) to [edge label = $W^-$] (p6);
            \propag [fermion] (p2) to [half left](p3);
            \propag [fermion] (p3) to [half left](p2);
            \propag [fermion] (p4) to [half left](p5);
            \propag [fermion] (p5) to [half left](p4);
            \vertex (p31) at (2.5,0.07);
            \vertex (p32) at (2.5,-0.07);
            \vertex (p41) at (3.5,0.07);
            \vertex (p42) at (3.5,-0.07);
            \propag [fermion] (p31) to [edge label = $\Pi_3$] (p41);
            \propag [fermion] (p42) to (p32);
        \end{feynhand}
        \end{tikzpicture}
    \]
    \[
        \begin{tikzpicture}
        \begin{feynhand}
            \vertex (p1) at (0,0);
            \vertex (p2) at (1,0);
            \vertex (p3) at (2.5,0);
            \vertex (p4) at (3.5,0);
            \vertex (p5) at (5,0);
            \vertex (p6) at (6,0);
            \propag [bos] (p1) to [edge label = $W^+$] (p2);
            \propag [bos] (p5) to [edge label = $W^+$] (p6);
            \propag [fermion] (p2) to [half left](p3);
            \propag [fermion] (p3) to [half left](p2);
            \propag [fermion] (p4) to [half left](p5);
            \propag [fermion] (p5) to [half left](p4);
            \vertex (p31) at (2.5,0.07);
            \vertex (p32) at (2.5,-0.07);
            \vertex (p41) at (3.5,0.07);
            \vertex (p42) at (3.5,-0.07);
            \propag [fermion] (p31) to [edge label = $\Pi_4$] (p41);
            \propag [fermion] (p42) to (p32);
        \end{feynhand}
        \end{tikzpicture}
    \]
    By calculating the four-points Green functions $G(\bar{f},f;\bar{f},f)$, we can get the amplitude of two quarks changing to the Goldstones because these Goldstones give the contributions around poles. 
    Recalling the Lagrangian in Eq.~(\ref{L1}), in the limit $\lambda_{ij}\rightarrow 0$ and $m_{ij}=0$, all fermions and gauge bosons are symmetric.
    So we conclude that their amplitudes translating to Goldstones are identical, and so do $F$'s, which have mass dimension and are defined by
    \begin{equation}
        p^\mu F=
        \begin{tikzpicture}[baseline=-0.08cm]
            \begin{feynhand}
                \vertex (p1) at (0,0);
                \vertex (p2) at (1,0);
                \vertex (p3) at (2.5,0);
                \vertex (p4) at (3.5,0);
                \vertex (p5) at (5,0);
                \vertex (p6) at (6,0);
                \propag [bos] (p1) to (p2);
                \propag [fermion] (p2) to [half left](p3);
                \propag [fermion] (p3) to [half left](p2);
                \vertex (p31) at (2.5,0.07);
                \vertex (p32) at (2.5,-0.07);
                \vertex (p41) at (3.5,0.07);
                \vertex (p42) at (3.5,-0.07);
                \propag [fermion] (p31) to [edge label = $\Pi$] (p41);
                \propag [fermion] (p42) to (p32);
            \end{feynhand}
            \end{tikzpicture}    
    \end{equation}
    Notice that $Z^0_\mu=\frac{1}{\sqrt{g^2+g'^2}}\left(gA^3_\mu-g'B_\mu\right)$, and $A_\mu=\frac{1}{\sqrt{g^2+g'^2}}\left(g'A^3_\mu+gB_\mu\right)$ as usual, where $A_\mu^a$ and $B_\mu$ are, respectively, the $\text{SU(2)}$ and $\text{U(1)}$ gauge bosons. In order to maintain $A_\mu$ massless, we should have $\left\langle A_\mu|\Pi\right\rangle=0$.
    If we write the mixing state of $\Pi_1$ and $\Pi_2$ which keeps massless and is eaten by $Z_0$ as $\pi_z$, while the other one as $h$, we have
    \begin{equation}
        \begin{aligned}
            &\left\langle A_\mu^3|\pi_z\right\rangle=gF,\\
            &\left\langle Z_\mu^0|\pi_z\right\rangle=-g'F.\\
        \end{aligned}
    \end{equation}
    By combining these aspects, the mass of $Z^0$ is given as
    \begin{eqnarray}
            m_Z^2&=&\frac{g}{\sqrt{g^2+g'^2}}\left(g\right)\left(g\right)\frac{g}{\sqrt{g^2+g'^2}}F^2\nonumber\\
            &&+\frac{g}{\sqrt{g^2+g'^2}}\left(g\right)\left(-g'\right)\frac{-g'}{\sqrt{g^2+g'^2}}F^2\nonumber\\
            &&+\frac{-g'}{\sqrt{g^2+g'^2}}\left(-g'\right)\left(g\right)\frac{g}{\sqrt{g^2+g'^2}}F^2\nonumber\\
            &&+\frac{-g'}{\sqrt{g^2+g'^2}}\left(-g'\right)\left(-g'\right)\frac{-g'}{\sqrt{g^2+g'^2}}F^2\nonumber\\
            &=&(g^2+g'^2)F^2.
    \end{eqnarray}
    Similarly, the mass of $A_\mu$ is given by
    \begin{eqnarray}
            m_A^2&=&\frac{g'}{\sqrt{g^2+g'^2}}\left(g\right)\left(g\right)\frac{g'}{\sqrt{g^2+g'^2}}F^2\nonumber\\
            &&+\frac{g'}{\sqrt{g^2+g'^2}}\left(g\right)\left(-g'\right)\frac{g}{\sqrt{g^2+g'^2}}F^2\nonumber\\
            &&+\frac{g}{\sqrt{g^2+g'^2}}\left(-g'\right)\left(g\right)\frac{g'}{\sqrt{g^2+g'^2}}F^2\nonumber\\
            &&+\frac{g}{\sqrt{g^2+g'^2}}\left(-g'\right)\left(-g'\right)\frac{g}{\sqrt{g^2+g'^2}}F^2\nonumber\\
            &=&0
    \end{eqnarray} as expected and
    \begin{equation}
        m_W^2=g^2F^2.
    \end{equation}
    Because $\cos\theta_w=\frac{g}{\sqrt{g^2+g'^2}}$, we see that $m_W=m_Z\cos\theta_w$.

    Because the other boson $h$ is a mixing of $\Pi_1$ and $\Pi_2$, it contains all components of $u_1$, $u_2$, $d_1$, and $d_2$. Therefore, $h$ has all coupling with fermions, with weak gauge bosons, and with itself.
    We can identify it with the Higgs boson in the Glashow-Weinberg-Salam model.

\section{SUMMARY}
    We studied the flavor change effects in a multi-flavor system.
    With the extended multi-flavor DS equations, we find that the flavor symmetry SU(2) is dynamically broken to SO(2). Combining the up-type quark sector and the down-type quark sector and take the Electroweak interaction as perturbation into consideration, we get the $\text{SU(4)}_L\rightarrow\text{SU(2)}_L\otimes\text{SU(2)}_L\rightarrow\text{SO(2)}_L\otimes \text{SO(2)}_L$ symmetry breaking chain. This would produce massless Goldstones, which can replace the role of the Higgs. Besides, due to the flavor mixing effects, the fermion mass eigenstates will misalign with the flavor eigenstates. By diagonalizing the mass matrix, the quark states are redefined, and the fermion masses are split apparently. We can interpret the new fermion mass eigenstates with distinct masses as different generations.

    We started with a simple point interaction model. By calculating this model analytically, we obtained the structure of their solutions. We find that there are always three sets of closely related solutions, which can be classified into three interrelated ellipses. These solutions generate three distinct masses, which can be identified as $m_u$, $m_c$, and $m_t$. So does the down quark sector.
    After that, we studied a more realistic interaction model. Although this model can only be solved numerically, we find the structures of solutions obtained before still hold.
    The masses of different generation quarks split obviously, and the constituent masses are always greater than their current masses. The higher generation quarks have almost the same constituent masses as the current masses, while for the low generation quarks, the constituent masses are much more significant due to dynamical effects.

    Only consider the left-handed part, the symmetry breaking produces fifteen Goldstones, but only four of them are independent. Two of them have electric charge $0$, and the other two have electric charge $+1$, $-1$, respectively. Among these Goldstones, one is pNGB and has a mass which could be identified as the Higgs boson. The other three Goldstones keep massless and will be eaten by gauge bosons to give $W^\pm$ and $Z_0$ masses via the Schwinger mechanism. Besides, the famous equation $m_W=m_Z\cos\theta_w$ is maintained.

    There is still much work to be done. The $m_Z$, $m_W$ and $m_H$ should be calculated exactly, and the Higgs couplings should be checked carefully. The related work is under progress.

\begin{acknowledgments}
    The work was supported by the National Natural Science Foundation of China under Contracts No. 11435001 and No. 11775041. Helpful discussions with Professors Shou-hua Zhu and Qing-hong Cao are acknowledged with great thanks.
\end{acknowledgments}

\appendix

\section{}
    The eight nonlinear integration equations for $\{V_i, S_i, T_i, R_i\}$ are as follows:
    \begin{eqnarray}
        1&=&\Big[Z_2^{11}Z_m^{11}m_{11}S_1\left(p^2\right)+Z_2^{11}p^2V_1\left(p^2\right)\nonumber\\
        &&+Z_m^{12}m_{12}R_2\left(p^2\right)+Z_\lambda^{12}\lambda_{12}p^2T_2\left(p^2\right)\Big]\nonumber\\
        &&+\int\frac{\mathrm{d}^4k}{\left(2\pi\right)^4}G(p-k)\bigg\{3\Big[S_1\left(k^2\right)S_1\left(p^2\right)\nonumber\\
        &&+R_1\left(k^2\right)R_2\left(p^2\right)\Big]+\Big[V_1\left(k^2\right)V_1\left(p^2\right)\nonumber\\
        &&+T_1\left(k^2\right)T_2\left(p^2\right)\Big]\Big[2p\cdot k\nonumber\\
        &&+\frac{\left(p^2+k^2\right)p\cdot k-2p^2k^2}{\left(p-k\right)^2}\Big]\bigg\},\\
        1&=&\Big[Z_m^{21}m_{21}R_1\left(p^2\right)+Z_\lambda^{21}\lambda_{21}p^2T_1\left(p^2\right)\nonumber\\
        &&+Z_2^{22}Z_m^{22}m_{22}S_2\left(p^2\right)+Z_2^{22}p^2V_2\left(p^2\right)\Big]\nonumber\\
        &&+\int\frac{\mathrm{d}^4k}{\left(2\pi\right)^4}G(p-k)\bigg\{3\Big[R_2\left(k^2\right)R_1\left(p^2\right)\nonumber\\
        &&+S_2\left(k^2\right)S_2\left(p^2\right)\Big]+\Big[T_2\left(k^2\right)T_1\left(p^2\right)\nonumber\\
        &&+V_2\left(k^2\right)V_2\left(p^2\right)\Big]\Big[2p\cdot k\nonumber\\
        &&+\frac{\left(p^2+k^2\right)p\cdot k-2p^2k^2}{\left(p-k\right)^2}\Big]\bigg\},
    \end{eqnarray}
    \begin{eqnarray}
        0&=&\Big[Z_2^{11}Z_m^{11}m_{11}R_1\left(p^2\right)+Z_2^{11}p^2T_1\left(p^2\right)\nonumber\\
        &&+Z_m^{12}m_{12}S_2\left(p^2\right)+Z_\lambda^{12}\lambda_{12}p^2V_2\left(p^2\right)\Big]\nonumber\\
        &&+\int\frac{\mathrm{d}^4k}{\left(2\pi\right)^4}G(p-k)\bigg\{3\Big[S_1\left(k^2\right)R_1\left(p^2\right)\nonumber\\
        &&+R_1\left(k^2\right)S_2\left(p^2\right)\Big]+\Big[V_1\left(k^2\right)T_1\left(p^2\right)\nonumber\\
        &&+T_1\left(k^2\right)V_2\left(p^2\right)\Big]\Big[2p\cdot k\nonumber\\
        &&+\frac{\left(p^2+k^2\right)p\cdot k-2p^2k^2}{\left(p-k\right)^2}\Big]\bigg\},\\
        0&=&\Big[Z_m^{21}m_{21}S_1\left(p^2\right)+Z_\lambda^{21}\lambda_{21}p^2V_1\left(p^2\right)\nonumber\\
        &&+Z_2^{22}Z_m^{22}m_{22}R_2\left(p^2\right)+Z_2^{22}p^2T_2\left(p^2\right)\Big]\nonumber\\
        &&+\int\frac{\mathrm{d}^4k}{\left(2\pi\right)^4}G(p-k)\bigg\{3\Big[R_2\left(k^2\right)S_1\left(p^2\right)\nonumber\\
        &&+S_2\left(k^2\right)R_2\left(p^2\right)\Big]+\Big[T_2\left(k^2\right)V_1\left(p^2\right)\nonumber\\
        &&+V_2\left(k^2\right)T_2\left(p^2\right)\Big]\Big[2p\cdot k\nonumber\\
        &&+\frac{\left(p^2+k^2\right)p\cdot k-2p^2k^2}{\left(p-k\right)^2}\Big]\bigg\},\\
        0&=&p^2\Big[Z_2^{11}S_1\left(p^2\right)-Z_2^{11}Z_m^{11}m_{11}V_1\left(p^2\right)\nonumber\\
        &&+Z_\lambda^{12}\lambda_{12}R_2\left(p^2\right)-Z_m^{12}m_{12}T_2\left(p^2\right)\Big]\nonumber\\
        &&+\int\frac{\mathrm{d}^4k}{\left(2\pi\right)^4}G(p-k)\bigg\{\!\!-\!3p^2\Big[S_1\left(k^2\right)V_1\left(p^2\right)\nonumber\\
        &&+R_1\left(k^2\right)T_2\left(p^2\right)\Big]+\Big[V_1\left(k^2\right)S_1\left(p^2\right)\nonumber\\
        &&+T_1\left(k^2\right)R_2\left(p^2\right)\Big]\Big[2p\cdot k\nonumber\\
        &&+\frac{\left(p^2+k^2\right)p\cdot k-2p^2k^2}{\left(p-k\right)^2}\Big]\bigg\},\\
        0&=&p^2\Big[Z_\lambda^{21}\lambda_{21}R_1\left(p^2\right)-Z_m^{21}m_{21}T_1\left(p^2\right)\nonumber\\
        &&+Z_2^{22}S_2\left(p^2\right)-Z_2^{22}Z_m^{22}m_{22}V_2\left(p^2\right)\Big]\nonumber\\
        &&+\int\frac{\mathrm{d}^4k}{\left(2\pi\right)^4}G(p-k)\bigg\{\!\!-\!3p^2\Big[R_2\left(k^2\right)T_1\left(p^2\right)\nonumber\\
        &&+S_2\left(k^2\right)V_2\left(p^2\right)\Big]+\Big[T_2\left(k^2\right)R_1\left(p^2\right)\nonumber\\
        &&+V_2\left(k^2\right)S_2\left(p^2\right)\Big]\Big[2p\cdot k\nonumber\\
        &&+\frac{\left(p^2+k^2\right)p\cdot k-2p^2k^2}{\left(p-k\right)^2}\Big]\bigg\},
    \end{eqnarray}
    \begin{eqnarray}
        0&=&p^2\Big[Z_2^{11}R_1\left(p^2\right)-Z_2^{11}Z_m^{11}m_{11}T_1\left(p^2\right)\nonumber\\
        &&+Z_\lambda^{12}\lambda_{12}S_2\left(p^2\right)-Z_m^{12}m_{12}V_2\left(p^2\right)\Big]\nonumber\\
        &&+\int\frac{\mathrm{d}^4k}{\left(2\pi\right)^4}G(p-k)\bigg\{\!\!-\!3p^2\Big[S_1\left(k^2\right)T_1\left(p^2\right)\nonumber\\
        &&+R_1\left(k^2\right)V_2\left(p^2\right)\Big]+\Big[V_1\left(k^2\right)R_1\left(p^2\right)\nonumber\\
        &&+T_1\left(k^2\right)S_2\left(p^2\right)\Big]\Big[2p\cdot k\nonumber\\
        &&+\frac{\left(p^2+k^2\right)p\cdot k-2p^2k^2}{\left(p-k\right)^2}\Big]\bigg\},\\
        0&=&p^2\Big[Z_\lambda^{21}\lambda_{21}S_1\left(p^2\right)-Z_m^{21}m_{21}V_1\left(p^2\right)\nonumber\\
        &&+Z_2^{22}R_2\left(p^2\right)-Z_2^{22}Z_m^{22}m_{22}T_2\left(p^2\right)\Big]\nonumber\\
        &&+\int\frac{\mathrm{d}^4k}{\left(2\pi\right)^4}G(p-k)\bigg\{\!\!-\!3p^2\Big[R_2\left(k^2\right)V_1\left(p^2\right)\nonumber\\
        &&+S_2\left(k^2\right)T_2\left(p^2\right)\Big]+\Big[T_2\left(k^2\right)S_1\left(p^2\right)\nonumber\\
        &&+V_2\left(k^2\right)R_2\left(p^2\right)\Big]\Big[2p\cdot k\nonumber\\
        &&+\frac{\left(p^2+k^2\right)p\cdot k-2p^2k^2}{\left(p-k\right)^2}\Big]\bigg\}.
    \end{eqnarray}
    
    Making use of $\{CV_{ij}, CS_{ij}, CT_{ij}, CR_{ij}\}$ ($i=1,2$), we obtain the simplified equations as follows:
    \begin{eqnarray}
        1&=&\Big[Z_2^{11}Z_m^{11}m_{11}S_1\left(p^2\right)+Z_2^{11}p^2V_1\left(p^2\right)\nonumber\\
        &&+Z_m^{12}m_{12}R_2\left(p^2\right)+Z_\lambda^{12}\lambda_{12}p^2T_2\left(p^2\right)\Big]\nonumber\\
        &&+\frac{3\pi}{2}CS_{11}S_1(p^2)+\frac{3\pi}{2}CR_{11}R_2(p^2)\nonumber\\
        &&-\frac{3\pi}{4}p^2CV_{11}V_1(p^2)+\frac{\pi}{4}p^4CV_{12}V_1(p^2)\nonumber\\
        &&-\frac{3\pi}{4}p^2CT_{11}T_2(p^2)+\frac{\pi}{4}p^4CT_{12}T_2(p^2),\qquad\quad
    \end{eqnarray}
    \begin{eqnarray}
        1&=&\Big[Z_m^{21}m_{21}R_1\left(p^2\right)+Z_\lambda^{21}\lambda_{21}p^2T_1\left(p^2\right)\nonumber\\
        &&+Z_2^{22}Z_m^{22}m_{22}S_2\left(p^2\right)+Z_2^{22}p^2V_2\left(p^2\right)\Big]\nonumber\\
        &&+\frac{3\pi}{2}CR_{21}R_1(p^2)+\frac{3\pi}{2}CS_{21}S_2(p^2)\nonumber\\
        &&-\frac{3\pi}{4}p^2CT_{21}T_1(p^2)+\frac{\pi}{4}p^4CT_{22}T_1(p^2)\nonumber\\
        &&-\frac{3\pi}{4}p^2CV_{21}V_2(p^2)+\frac{\pi}{4}p^4CV_{22}V_2(p^2),\\
        0&=&\Big[Z_2^{11}Z_m^{11}m_{11}R_1\left(p^2\right)+Z_2^{11}p^2T_1\left(p^2\right)\nonumber\\
        &&+Z_m^{12}m_{12}S_2\left(p^2\right)+Z_\lambda^{12}\lambda_{12}p^2V_2\left(p^2\right)\Big]\nonumber\\
        &&+\frac{3\pi}{2}CS_{11}R_1(p^2)+\frac{3\pi}{2}CR_{11}S_2(p^2)\nonumber\\
        &&-\frac{3\pi}{4}p^2CV_{11}T_1(p^2)+\frac{\pi}{4}p^4CV_{12}T_1(p^2)\nonumber\\
        &&-\frac{3\pi}{4}p^2CT_{11}V_2(p^2)+\frac{\pi}{4}p^4CT_{12}V_2(p^2),\\
        0&=&\Big[Z_m^{21}m_{21}S_1\left(p^2\right)+Z_\lambda^{21}\lambda_{21}p^2V_1\left(p^2\right)\nonumber\\
        &&+Z_2^{22}Z_m^{22}m_{22}R_2\left(p^2\right)+Z_2^{22}p^2T_2\left(p^2\right)\Big]\nonumber\\
        &&+\frac{3\pi}{2}CR_{21}S_1(p^2)+\frac{3\pi}{2}CS_{21}R_2(p^2)\nonumber\\
        &&-\frac{3\pi}{4}p^2CT_{21}V_1(p^2)+\frac{\pi}{4}p^4CT_{22}V_1(p^2)\nonumber\\
        &&-\frac{3\pi}{4}p^2CV_{21}T_2(p^2)+\frac{\pi}{4}p^4CV_{22}T_2(p^2),\\
        0&=&\Big[Z_2^{11}S_1\left(p^2\right)-Z_2^{11}Z_m^{11}m_{11}V_1\left(p^2\right)\nonumber\\
        &&+Z_\lambda^{12}\lambda_{12}R_2\left(p^2\right)-Z_m^{12}m_{12}T_2\left(p^2\right)\Big]\nonumber\\
        &&-\frac{3\pi}{2}CS_{11}V_1(p^2)-\frac{3\pi}{2}CR_{11}T_2(p^2)\nonumber\\
        &&-\frac{3\pi}{4}CV_{11}S_1(p^2)+\frac{\pi}{4}p^2CV_{12}S_1(p^2)\nonumber\\
        &&-\frac{3\pi}{4}CT_{11}R_2(p^2)+\frac{\pi}{4}p^2CT_{12}R_2(p^2),\\
        0&=&\Big[Z_\lambda^{21}\lambda_{21}R_1\left(p^2\right)-Z_m^{21}m_{21}T_1\left(p^2\right)\nonumber\\
        &&+Z_2^{22}S_2\left(p^2\right)-Z_2^{22}Z_m^{22}m_{22}V_2\left(p^2\right)\Big]\nonumber\\
        &&-\frac{3\pi}{2}CR_{21}T_1(p^2)-\frac{3\pi}{2}CS_{21}V_2(p^2)\nonumber\\
        &&-\frac{3\pi}{4}CT_{21}R_1(p^2)+\frac{\pi}{4}p^2CT_{22}R_1(p^2)\nonumber\\
        &&-\frac{3\pi}{4}CV_{21}S_2(p^2)+\frac{\pi}{4}p^2CV_{22}S_2(p^2),\qquad\qquad
    \end{eqnarray}
    \begin{eqnarray}
        0&=&\Big[Z_2^{11}R_1\left(p^2\right)-Z_2^{11}Z_m^{11}m_{11}T_1\left(p^2\right)\nonumber\\
        &&+Z_\lambda^{12}\lambda_{12}S_2\left(p^2\right)-Z_m^{12}m_{12}V_2\left(p^2\right)\Big]\nonumber\\
        &&-\frac{3\pi}{2}CS_{11}T_1(p^2)-\frac{3\pi}{2}CR_{11}V_2(p^2)\nonumber\\
        &&-\frac{3\pi}{4}CV_{11}R_1(p^2)+\frac{\pi}{4}p^2CV_{12}R_1(p^2)\nonumber\\
        &&-\frac{3\pi}{4}CT_{11}S_2(p^2)+\frac{\pi}{4}p^2CT_{12}S_2(p^2),\\
        0&=&\Big[Z_\lambda^{21}\lambda_{21}S_1\left(p^2\right)-Z_m^{21}m_{21}V_1\left(p^2\right)\nonumber\\
        &&+Z_2^{22}R_2\left(p^2\right)-Z_2^{22}Z_m^{22}m_{22}T_2\left(p^2\right)\Big]\nonumber\\
        &&-\frac{3\pi}{2}CR_{21}V_1(p^2)-\frac{3\pi}{2}CS_{21}T_2(p^2)\nonumber\\
        &&-\frac{3\pi}{4}CT_{21}S_1(p^2)+\frac{\pi}{4}p^2CT_{22}S_1(p^2)\nonumber\\
        &&-\frac{3\pi}{4}CV_{21}R_2(p^2)+\frac{\pi}{4}p^2CV_{22}R_2(p^2).\qquad\quad
    \end{eqnarray}

\section{}
    We write the generators of $SU(4)$ as $\{T_a\}$, which satisfies
    \begin{equation}
        \textrm{Tr} T_aT_b=2\delta_{ab}.
    \end{equation}
    The structure constants of the Lie algebra are
    \begin{equation}
        \left[T_a,T_b\right]=\sum_cif_{abc}T_c.
    \end{equation}
    Moreover, we have 
    \begin{equation}
        \left[Q_a,\psi\right]=-T_a\psi
    \end{equation}
    where
    \begin{equation}
        Q_a=\int \mathrm{d}\textbf{r}\psi^\dagger T_a\psi
    \end{equation}
    is the charge corresponding to $T_a$.
    We set M to be the superposition of two bases: $M=T_m=T_1+T_{13}$.

    We begin with
    \begin{eqnarray}\label{B1}
        &&e^{i\theta Q_a}\bar{\psi}e^{-i\theta Q_a}\left[M,T_a\right]e^{i\theta Q_a}\psi e^{-i\theta Q_a}\nonumber\\
        &&=\left(\bar{\psi}+i\theta\bar{\psi} T_a\right)\left[M,T_a\right]\left(\psi-i\theta T_a\psi\right)\nonumber\\
        &&=\bar{\psi}\left[M,T_a\right]\psi+i\theta \bar{\psi}\left[T_a,\left[M,T_a\right]\right]\psi+O(\theta^2)\nonumber\\
        &&=-i\sum_c f_{amc}\bar{\psi}T_c\psi-i\theta\sum_{cd}f_{acm}f_{acd}\bar{\psi}T_d\psi\nonumber\\
        &&\quad+O\left(\theta^2\right).
    \end{eqnarray}
    Notice that from here, $m$ is an index needed to sum for $m=1$ and $m=13$, for instance, $f_{amc}=f_{a1c}+f_{a13c}$.
    We assume that $\langle\bar{\psi}M\psi\rangle=\langle\bar{\psi}T_m\psi\rangle\neq0$, that is, $\langle\bar{\psi}T_1\psi\rangle\neq0$, $\langle\bar{\psi}T_{13}\psi\rangle\neq0$, but $\langle\bar{\psi}T_d\psi\rangle=0$ when $d\neq1$ or $d\neq13$.
    Then, we define
    \begin{equation}
        \Delta=\Delta_1+\Delta_2,
    \end{equation}
    where
    \begin{eqnarray}
        \Delta_1=&&\langle\bar{\psi}T_1\psi\rangle,\\
        \Delta_2=&&\langle\bar{\psi}T_{13}\psi\rangle.
    \end{eqnarray} 
    Notice that $f_{1a13}=0$, we have
    \begin{eqnarray}
        \sum_c f_{amc}\langle\bar{\psi}T_c\psi\rangle=&&f_{am1}\langle\bar{\psi}T_1\psi\rangle+f_{am13}\langle\bar{\psi}T_{13}\psi\rangle\nonumber\\
        =&&f_{a,1,1}\langle\bar{\psi}T_1\psi\rangle+f_{a,13,1}\langle\bar{\psi}T_1\psi\rangle\nonumber\\
        &&+f_{a,1,13}\langle\bar{\psi}T_{13}\psi\rangle+f_{a,13,13}\langle\bar{\psi}T_{13}\psi\rangle\nonumber\\
        =&&0.
    \end{eqnarray}
    Also, because of $\sum_cf_{ac1}f_{ac13}=0$, we have
    \begin{eqnarray}
        &&\sum_{cd}f_{acm}f_{acd}\bar{\psi}T_d\psi\nonumber\\
        =&&\sum_{cd}f_{ac1}f_{acd}\bar{\psi}T_d\psi+\sum_{cd}f_{ac13}f_{acd}\bar{\psi}T_d\psi\nonumber\\
        =&&\sum_cf_{ac1}f_{ac1}\bar{\psi}T_1\psi+\sum_cf_{ac1}f_{ac13}\bar{\psi}T_{13}\psi\nonumber\\
        &&+\sum_cf_{ac13}f_{ac1}\bar{\psi}T_1\psi+\sum_cf_{ac13}f_{ac13}\bar{\psi}T_{13}\psi\nonumber\\
        =&&\sum_cf_{ac1}^2\bar{\psi}T_1\psi+\sum_cf_{ac13}^2\bar{\psi}T_{13}\psi\nonumber\\
        =&&\sum_cf_{ac1}^2\Delta_1+\sum_cf_{ac13}^2\Delta_2.
    \end{eqnarray}
    These result in
    \begin{eqnarray}\label{B2}
        &&\sum_c f_{ac1}^2 \Delta_1+\sum_c f_{ac13}^2 \Delta_2\nonumber\\
        &&=\lim_{\theta\to 0}\frac{-1}{i\theta}e^{i\theta Q_q}\bar{\psi}e^{-i\theta Q_a}\left[M,T_a\right]e^{i\theta Q_a}\psi e^{-i\theta Q_a}\nonumber\\
        &&=\lim_{\theta\to 0}\frac{-1}{\theta}e^{i\theta Q_a}\Pi_a e^{-i\theta Q_a};
    \end{eqnarray}
    in the third line, we have made use of $e^{-i\theta Q_a}\left[M,T_a\right]e^{i\theta Q_a}=\left[M,T_a\right]$ because $Q_a$ is an operator (not matrix).
    
    We add the symmetry breaking term $H_\lambda$ to the original Hamiltonian $H_0$ and write the total Hamiltonian as
    \begin{equation}
        H_T=H_0+H_\lambda,
    \end{equation}
    where
    \begin{equation}
        H_\lambda=-\lambda\int\mathrm{d}\mathbf{r}\bar{\psi}T_1\psi-\lambda\int\mathrm{d}\mathbf{r}\bar{\psi}T_{13}\psi.
    \end{equation}
    Furthermore, we write $\phi$ as the ground state of the total Hamiltonian $H_T$, that is $H_T\phi=E_0\phi$.
    We define
    \begin{eqnarray}
        A_a&&\equiv\langle\phi|e^{i\theta Q_a}\Pi_a(\textbf{r})e^{-i\theta Q_a}|\phi\rangle\nonumber\\
        &&=\langle\phi|e^{iH_Tt}e^{i\theta Q_a}\Pi_a(\textbf{r})e^{-i\theta Q_a}e^{-iH_Tt}|\phi\rangle.\nonumber\\
    \end{eqnarray}
    Following Ref.~\cite{NGBCBTOPISSB}, we get the same result:
    \begin{eqnarray}\label{B3}
        A_a=&&\langle\phi|\Pi_a|\phi\rangle\nonumber\\
        &&-\theta\lambda\int\mathrm{d}\textbf{r}'\int_{-\infty}^\infty\mathrm{d}t'D_{aa}^R\left(t-t',\textbf{r}-\textbf{r}'\right)\nonumber\\
        &&+O(H_I^2),
    \end{eqnarray}
    where the retarded Green's function is defined as
    \begin{equation}
        D_{aa}^R(t-t',\textbf{r}-\textbf{r}')\!=\!-i\theta(t-t')\langle\left[\Pi_a(\textbf{r},t),\Pi_a(\textbf{r}'\!,t')\right]\rangle,
    \end{equation}
    and
    \begin{equation}
        H_I=-\theta\lambda\int\mathrm{d}\textbf{r}\Pi_a(\textbf{r},t).
    \end{equation}
    Substituting Eq.~(\ref{B3}) and $\langle\phi|\Pi_a|\phi\rangle=0$ to Eq.~(\ref{B2}), and taking the Fourier transformation and analytic continuation \cite{abrikosov1965quantum}, we have
    \begin{equation}
        \sum_cf_{ac1}^2\Delta_1+\sum_cf_{ac13}^2\Delta_2=\lambda D_{aa}(\omega=0,\textbf{q}=0).
    \end{equation}

\bibliography{bibli}
\end{document}